\documentclass[aps,pra,showpacs,notitlepage,floatfix,
twocolumn,superscriptaddress]{revtex4-1}
\usepackage{bbm}
\usepackage{mathrsfs}
\usepackage{epsfig}
\usepackage{graphicx}
\usepackage{amsfonts}
\usepackage[figuresright]{rotating}
\usepackage{amssymb}
\usepackage{amsmath}
\usepackage{dcolumn}
\usepackage{bm}
\usepackage{braket}
\DeclareMathOperator{\Tr}{Tr}
\usepackage[colorlinks,linkcolor=blue,anchorcolor=blue,citecolor=blue,urlcolor=blue]{hyperref}

\def\Re{\rm{Re}}
\def\Im{\rm{Im}}
\def\be{\begin{equation}} \def\ee{\end{equation}}
\def\bea{\begin{equation}\begin{aligned}} \def\eea{\end{aligned}\end{equation}}

\begin{document}
\title{Accessing scrambling using matrix product operators}

\author{Shenglong Xu}
\affiliation{Condensed Matter Theory Center and Department of Physics, University of Maryland, College Park, MD 20742, USA}
\author{Brian Swingle}
\affiliation{Condensed Matter Theory Center, Maryland Center for Fundamental Physics, Joint Center for Quantum Information and Computer Science, and Department of Physics, University of Maryland, College Park, MD 20742, USA}

\begin{abstract}
Scrambling, a process in which quantum information spreads over a complex quantum system becoming inaccessible to simple probes, happens in generic chaotic quantum many-body systems, ranging from spin chains, to metals, even to black holes. Scrambling can be measured using out-of-time-ordered correlators (OTOCs), which are closely tied to the growth of Heisenberg operators. In this work, we present a general method to calculate OTOCs of local operators in local one-dimensional systems based on approximating Heisenberg operators as matrix-product operators (MPOs). Contrary to the common belief that such tensor network methods work only at early times, we show that the entire early growth region of the OTOC can be captured using an MPO approximation with modest bond dimension. We analytically establish the goodness of the approximation by showing that if an appropriate OTOC is close to its initial value, then the associated Heisenberg operator has low entanglement across a given cut. We use the method to study scrambling in a chaotic spin chain with $201$ sites. Based on this data and OTOC results for black holes, local random circuit models, and non-interacting systems, we conjecture a universal form for the dynamics of the OTOC near the wavefront. We show that this form collapses the chaotic spin chain data over more than fifteen orders of magnitude.
\end{abstract}

\maketitle
\textit{Introduction---}Scrambling is a process of spreading and effective loss of quantum information in complex quantum many-body system, and it has received considerable recent attention, going back to work on the black hole information problem~\cite{ Hayden2007, Sekino2008, Shenker2014, Hosur2016}. One relevance of scrambling is to the problem of quantum thermalization (e,g.,~\cite{Deutsch1991, Srednicki1994,Tasaki1998a,Rigol2008}). Indeed, the process of ``hiding'' information behind a black hole horizon shares many similarities to the effective loss of information in quantum thermalization. Scrambling describes the emergence of irreversibility in unitary dynamics, when the information about the initial condition spreads and can no longer be inferred from local measurements of few-body operators. In the context of spatially extended condensed matter systems and other quantum many-body systems, relaxation to local equilibrium is a relatively fast process while scrambling takes longer, with a time-scale depending on system size. Quantum field theories with AdS/CFT dual descriptions involving black holes \cite{Shenker2014} along with the Sachdev-Ye-Kitaev (SYK) model \cite{Sachdev1993,kitaev2015} and coupled SYK clusters \cite{Gu2017} exhibit the fastest possible scrambling rate~\cite{Maldacena2016}, while the scrambling in many-body localized (MBL) system is much slower \cite{Swingle2016a, Chen2016a, Fan2017, Huang2017}. With the rapid advancement of experimental techniques, measurements of scrambling \cite{swingle2016,Zhu2016,Yao2016a,Halpern2016,Halpern2017,Campisi2017,Yoshida2017,Garttner2016, Wei2016, Li2017a,  Meier2017} may soon become a standard probe, in addition to response functions, to characterize quantum dynamics in ultra-cold atom systems. In short, scrambling is an importance concept lying at the interface between quantum information, quantum matter and quantum gravity.

One measure of scrambling is the growth of local Heisenberg operators. This growth can be characterized by the time-dependence of the expectation value of the squared commutator of two local operators, $W$ and $V$, as function of their separation \cite{Shenker2014},
\be
C(t)=\braket{[W(t), V]^\dagger [W(t), V]},
\label{eq:commutator}
\ee
where $W(t) = e^{i H t} W e^{-i H t}$ is a Heisenberg operator. This squared commutator is directly related to the so-called out-of-time-ordered correlator (OTOC),
\be
F(t)=\braket{W(t)^\dagger V^\dagger W(t)V},
\ee
through the relation $C(t)=2(1-{\Re}[ F(t)])$, valid for unitary $W$ and $V$. Several related measures \cite{Leviatan2017, Chen2017, Bordia2018} have also been investigated. The OTOC has been extensively investigated across a wide variety of systems, including quantum field theories \cite{Roberts2015, Stanford2016, Chowdhury2017, Patel2017}, random unitary models~\cite{Nahum2017, Nahum2017a, Khemani2017, Rakovszky2017}, as well as spin chains \cite{Luitz2017, Bohrdt2017a, Heyl2018, Lin2018}. For systems with a geometrically local Hamiltonian which are not in a localized phase, $C(x, t)$ typically exhibits a ballistic traveling wavefront, with rapid growth ahead of the wavefront and saturation behind the wavefront at late time. In energy-conserving quantum chaotic systems at temperature $T$, the growth rate is bounded by $2\pi T$ \cite{Maldacena2016} and is proposed to diagnose quantum chaos by providing an analog of a classical Lyapunov exponent.

However, a systematic understanding of the specific growth form of $C(t)$ still remains elusive, mainly due to the lack of a controlled method for calculating OTOCs for generic local Hamiltonians at large system size. Here we introduce such a method, valid for one-dimensional systems, based on representing Heisenberg operators as matrix product operators (MPOs). Matrix product state/operator based methods, such as TEBD \cite{Vidal2006}, while excellent for simulating ground states of large systems, typically become incapable of capturing real-time dynamics beyond early time due to the rapid accumulation of entanglement. One of our main contributions is showing how this limitation can be circumvented to study the growth behavior of OTOCs even at late time. Recently, two modified matrix product state/operator methods \cite{Leviatan2017, White2017} were proposed, with both designed to respect conservation laws and produce late-time hydrodynamic behavior in response functions. However, calculating OTOCs with these methods remains challenging, and the effect of truncation error is unclear.

We use our method to study the early growth region of scrambling, in particular its behavior around the wavefront, in a chaotic spin chain with $201$ sites. We present two main results. First, motivated by existing results on black holes and local random circuit models as well as results on non-interacting systems derived in this paper, we conjecture a ``universal'' form for the growth of commutators,
\be
C(x,t)\sim \exp\left(-\lambda \frac{(x-v_B t)^{1+p}}{t^p}\right),
\label{eq:fitting_form}
\ee
where $x$ is the distance between $W$ and $V$, $v_B$ is the speed of the wavefront, and $p$ controls the spreading of the wavefront. Second, based on the close relation between operator spreading and scrambling, we establish that MPO dynamics can be used to capture the entire early growth behavior of the scrambling even with small bond dimension. Truncation only leads to errors in the operator representation behind the wavefront, but since such errors must still propagate outwards, the physics ahead of and up to the wavefront is captured accurately. We find excellent agreement between our analytical arguments and the numerical results for various models.

\begin{figure}
\centering
\includegraphics[height=0.45\columnwidth, width=0.45\columnwidth]
{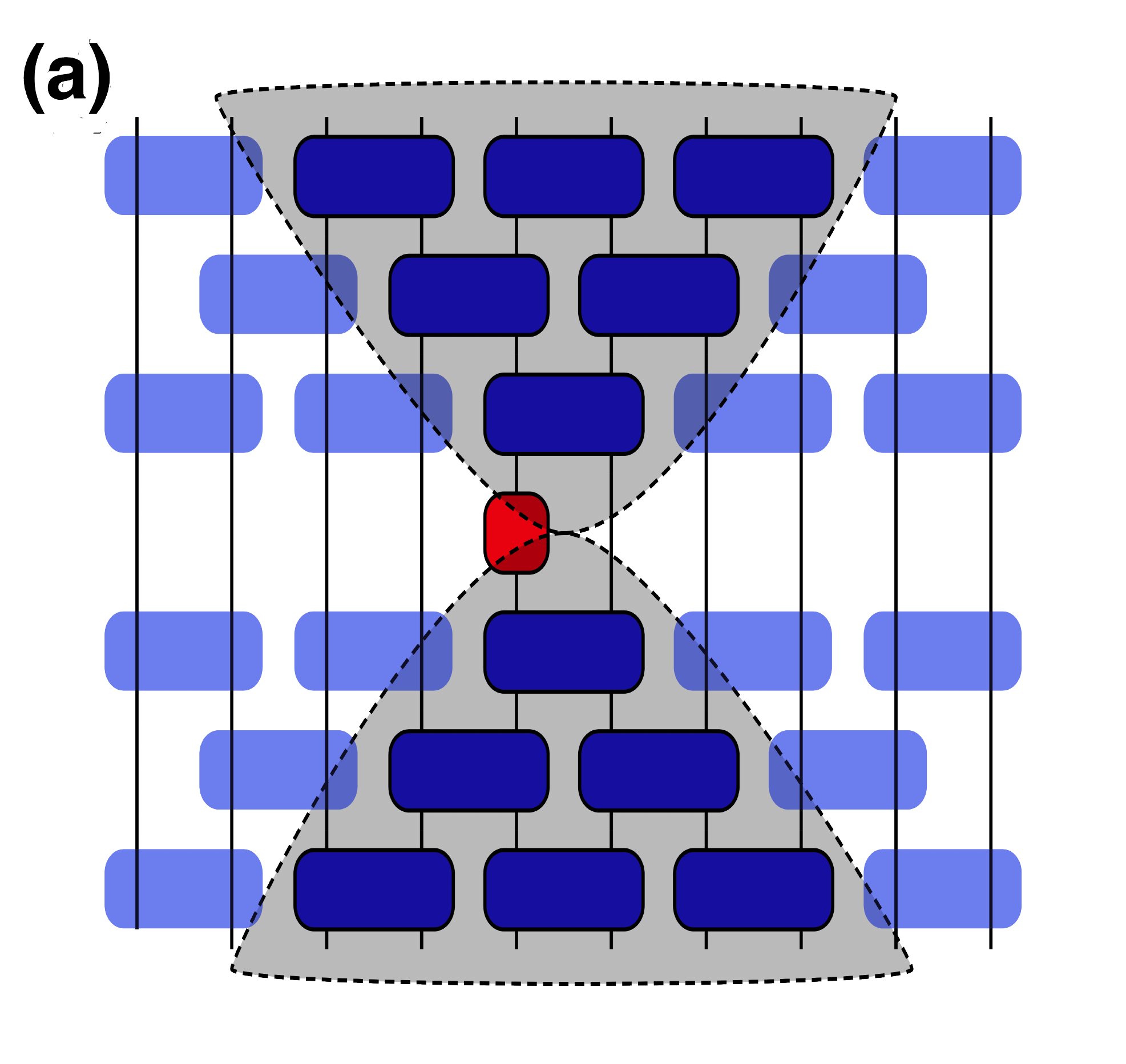}
\includegraphics[height=0.45\columnwidth, width=0.45\columnwidth]
{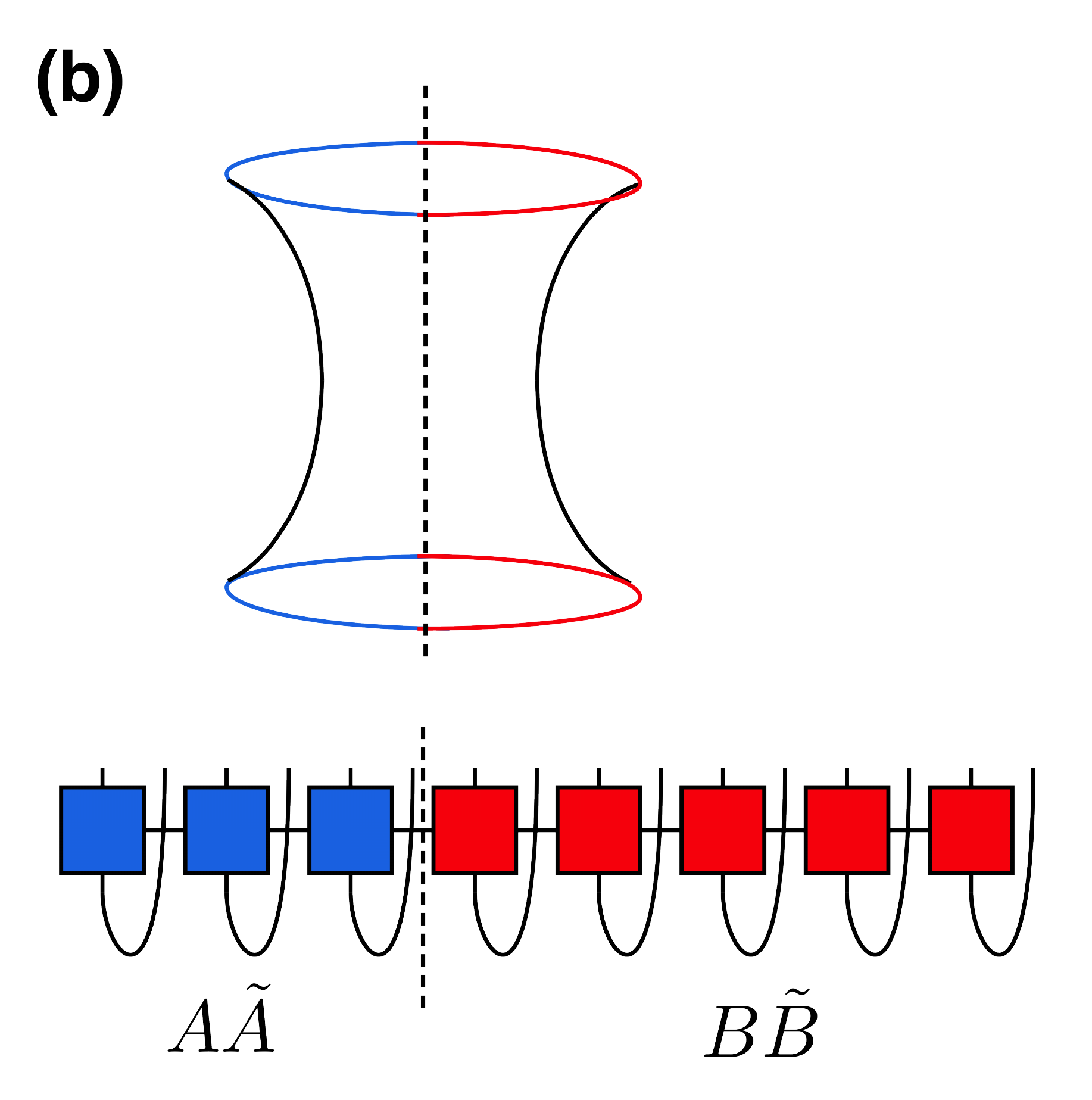}
\caption{(a) Light cone structure of a time-evolving ($U$) MPO, $U W U^\dagger$. The MPO starts as a local operator (red), and the initial required bond dimension is one. Under time evolution, higher bond dimension is only required in the shaded light cone region, since $U$ and $U^\dagger$ cancel outside the shaded region. Even inside the exact light cone, $U$ and $U^\dagger$ will approximately cancel for some time, so that the needed bond dimension remains small. (b) Entanglement of the MPO defined by treating it as a state with double the number of physical degrees. More generally, we can apply the MPO to a purified thermal state at temperature $T$ (the ``thermofield double'') and define a temperature dependent notion of operator entanglement. For systems with a holographic gravity dual (AdS/CFT), this temperature dependent operator entanglement is related to the entanglement between two halves of a two-sided wormhole. We show in Appendix~\ref{appsc:operator_entanglement} that our results on operator entanglement are also valid in holographic systems at finite temperature.}
\label{fig:tensor}
\end{figure}

\textit{Growth form---}There have been many recent proposals for the form of the growth of the commutator near the wavefront. A detailed discussion of the various forms can be found in Appendix~\ref{appsc:growth_form}. In particular, as reviewed above, results exist for holographic models and coupled SYK clusters at large $N$ (large number of local degrees of freedom), for local random circuit models, at weak coupling in various field theory models, and beyond. In holographic \cite{Shenker2014} and coupled SYK cluster models \cite{Gu2017,Song2017,Ben-Zion2017}, one finds $C \sim e^{\lambda (t-x/v_B)}$, where $\lambda$ called the quantum Lyapunov exponent. Various weak coupling calculations using resummed perturbation theory also give exponential growth, at least near the wavefront \cite{Chowdhury2017,Patel2017}. By contrast, in local random circuit models one finds $C(x,t)\sim e^{-\lambda (x-v_B t)^2/t}$~\cite{Nahum2017, Nahum2017a, Khemani2017, Rakovszky2017}, implying a diffusively spreading wavefront in contrast to the sharp wavefront above. In Appendix~\ref{appsc:noninteracting}, we show analytically that for non-interacting systems with translational symmetry, $C(x,t)\sim e^{-\lambda (x-v_B t)^{3/2}/t^{1/2}}$ where $v_B$ is the maximal group velocity and $\lambda$ some dispersion-dependent parameter. Motivated by these results, we conjecture a universal growth form for the growth of $C$ in generic many-body systems around the wavefront, Eq.~\eqref{eq:fitting_form}.

To further motivate this form, we observe that is it uniquely determined if we assume a ballistically expanding wavefront, finite growth rate of $C(x,t)$ near the wavefront (meaning $x$ and $t$ large, with $x/t$ finite and close $v_B$), and the Lieb-Robinson bound~\cite{Lieb1972,Nachtergaele2010}. The slow scrambling that occurs in many-body localized systems is excluded because of the logarithmic, rather than ballistic, expansion of wavefront in that case~\cite{Swingle2016a, Chen2016a, Fan2017, Huang2017}. The results we reviewed above can all be fit into this universal framework, with $p=0$ corresponding to pure exponential growth (holographic models, coupling SYK clusters, etc.), with $p=1/2$ for non-interacting particle models, and with $p=1$ for the local random circuit model. Therefore, it is very interesting to study $C(x,t)$ in other systems and compare the early growth behavior with Eq.~\eqref{eq:fitting_form}. However, current analytical results are often limited to weak coupling and/or large $N$ and reliable numerical results only exist for small systems. To address these challenges, we developed a numerical method to calculate the early growth dynamics of OTOCs in generic geometrically local many-body Hamiltonians based on the entanglement structure of local Heisenberg operators.

\textit{Operator entanglement and OTOC---}The operator entanglement of a generic operator $W$ in the Hilbert space can be obtained by defining an operator state as follows. Begin with the ``thermofield double'' state at infinite temperature, which is the maximally entangled state between the original system (sites $r$) and its partner (sites $\tilde{r}$),
\be
\ket{\Phi} = \prod_{r=1}^n \frac{\ket{00}_{r \tilde{r}}+\ket{11}_{r\tilde{r}}}{\sqrt{2}},
\ee
where $\ket{a}_r$ are eigenstates of Pauli $Z$: $Z_r \ket{a}_r = (-1)^a \ket{a}_r$. Then the operator state $\ket{W}$ is obtained by acting $W \otimes I$ on $\Phi$,
\be
\ket{W}=W\otimes I \ket{\Phi}.
\ee
The operator entanglement of a region $B$ and its partner $\tilde{B}$, denoted $S(B\tilde{B})$, is the von Neumann entropy of the $B\tilde{B}$ density matrix in total state $\ket{W}$. This object can be obtained as usual by tracing out the complementary regions $A$ and $\tilde{A}$ from the state $\ket{W}$. Intuitively, the operator entanglement entropy determines the minimal bond dimension required for an MPO to well approximate $W$.

We first analyze the dynamical entanglement structure of a local operator, say $X$, under Heisenberg time evolution, $X(t)=e^{iHt}Xe^{-iHt}$. The tensor network picture of the Heisenberg dynamics is sketched in Fig.~\ref{fig:tensor}(a). The operator takes a simple product form at $t=0$, and gradually builds up complexity as $t$ increases. The Lieb-Robinson bound approximately confines the expansion of the operator to a light cone; At least in chaotic systems with local Hamiltonians, local Heisenberg operators are expected to expand ballistically with velocity $v_B$; the Lieb-Robinson bound requires $v_B$ to the less than or equal to the Lieb-Robinson velocity. The tensor network representation naturally adapts to such light cone structure. As seen from Fig.~\ref{fig:tensor}(a), the unitaries outside the shaded light cone region approximately cancel each other, and that part of $X(t)$ remains identity. Consequently, if we partition the system into the set of all sites within $x$ of the location of $X$ and the complement, the operator entanglement of the partition, $S(x,t)$, will be small for $t \lesssim x/v_B$ and will grow linearly with $t$ for $t > x/v_B$ until it saturates to a value proportional to $x$ (``thermalization'' of the operator state).

When an MPO with fixed bond-dimension $\chi$ is used to represent $X(t)$, we expect that the truncation will introduce substantial error at bond $x$ once the operator entanglement reaches $S(x,t) \sim \log \chi$. Crucially, such truncation errors are only significant within the light cone, and these errors, regarded as a local perturbation to the operator, also propagate ballistically and do not escape the light cone. This is in sharp contrast with usual matrix product state (MPS) approaches to dynamics, where the analog of $S(x,t)$ grows linearly with time from $t=0$ and significant errors from truncation occur all throughout the system after a short time $\delta t$ defined by $S(x,\delta t) \sim \log \chi$. The MPO approach gains us access to a much wider space-time region $t<x/v_B+\delta t$, which we call the early-growth region, as compared to the early-time region $t<\delta t$ in the MPS approach.

Physically, we expect that the operator entanglement of $X(t)$ is closely related to the commutator (and OTOC) of $X(t)$ with local operators. To make the connection rigorous, it is convenient to assume $X$ is unitary. Then the results of Ref.~\cite{Hosur2016} can be extended by replacing the time evolution with the ``Heisenberg time evolution'', $e^{-i H t} \rightarrow e^{i H t} X e^{-i H t}$. From this starting point, we prove in Appendix~\ref{appsc:operator_entanglement} that the second R\'enyi entropy of the operator, $S^{(2)}(r,t)$, is bounded by the OTOC:
\be
S^{(2)}(B\tilde{B})\leq -\log\left(1-\frac{1}{2}\sum_{r\in B, O_r} ||[X(t), O_r]||^2\right)
\label{eq:bound}
\ee
where $X(t)$ at $t=0$ resides in the region A, and the summation is over all four \textit{local} Pauli operators $O_r \in\{I, X, Y, Z\}$ defined on site $r$. When the squared commutators are small, i.e, in the early-growth region, we expect each term in the sum to decay exponentially (with exponent $a$) or faster, and therefore the sum is dominated by the terms in which $O_r$ has support on the boundary between $A\tilde{A}$ and $B\tilde{B}$. After Taylor expanding Eq.~ \eqref{eq:bound} with these assumptions, we get
\be
S^{(2)}(B\tilde{B})\leq  \frac{1}{2(1-e^{-2a})}\sum\limits_{O_i}||[X(t), O_i]||^2
\ee
where again $a$ exponential decay constant (for example, in the Lieb-Robinson bound). Since the prefactor is order unity, this result proves that small $C(x,t)$ directly leads to small entanglement with the cut at $x$. Technically, the smallness of the second R\'enyi entropy does not guarantee a good MPO approximation, but in the physical cases of interest we expect that it is representative. As a result, truncation of the MPO along a bond at position $x$ is required only when $C(x,t)$ is significantly larger than zero, i.e., outside the early growth region. We also know that truncation error outside the early growth region (within the light cone) does not escape. These two pieces of information guarantee that an MPO with modest bond dimension can capture the entire early growth region.

\textit{Models---}As discussed in Appendices~\ref{appsc:growth_form} and \ref{appsc:noninteracting}, the early growth region, where Eq.~\eqref{eq:fitting_form} potentially applies, is distinct from the early time growth of $\frac{t^x}{x!}$, which is simply the first non-vanishing pertubative contribution to $C(x,t)$. We applied the above MPO approach to various models to calculate the early growth of the commutator and find excellent convergence of the results with the bond dimension, as expected from the reasoning above. More interestingly, all the models exhibit early growth behavior that is consistent with Eq.~\eqref{eq:fitting_form}. For concreteness, we study the following mixed-field quantum Ising model in one dimension,
\be
H=-\frac{1}{E_0}\left ( J \sum\limits_{r=1}^{L} Z_r Z_{r+1} +h_x\sum\limits_{r=1}^L X_r+h_z\sum\limits_{r=1}^L Z_r \right)
\label{eq:ham}
\ee
where $Z_r$ and $X_r$ are local Pauli operators. We choose the normalization factor $E_0=\sqrt{4J^2+2h_x^2+2h_z^2}$ so that the Hamiltonian defined on a bond squares to $1$.  When the longitudinal field $h_z$ is zero, this is the transverse field Ising model which has a non-interacting fermion description. As $h_z$ is turned on, the model is generally non-integrable and has been widely used to study quantum chaos and thermalization. In the following, we study this model in both the non-interacting region and the chaotic region, with the parameters $J=1$, $h_x=1.05$, and $h_z=0$ or $0.5$ representing the two different regions, respectively. The system size is set to $L=201$ spins except when stated otherwise. Benchmarking with exact diagonalization results on smaller system sizes is shown in the supplementary material. In practice, we use a DMRG-like sweeping algorithm, slightly different from Fig.~\ref{fig:tensor}(a). We are aiming at intermediate time scales of order $t \sim 150$ (in units where $J=1$), and a small Trotter step $dt = 0.005$ is applied to reduce the error from the time discretization.

\begin{figure}
\includegraphics[height=0.45\columnwidth, width=0.45\columnwidth]
{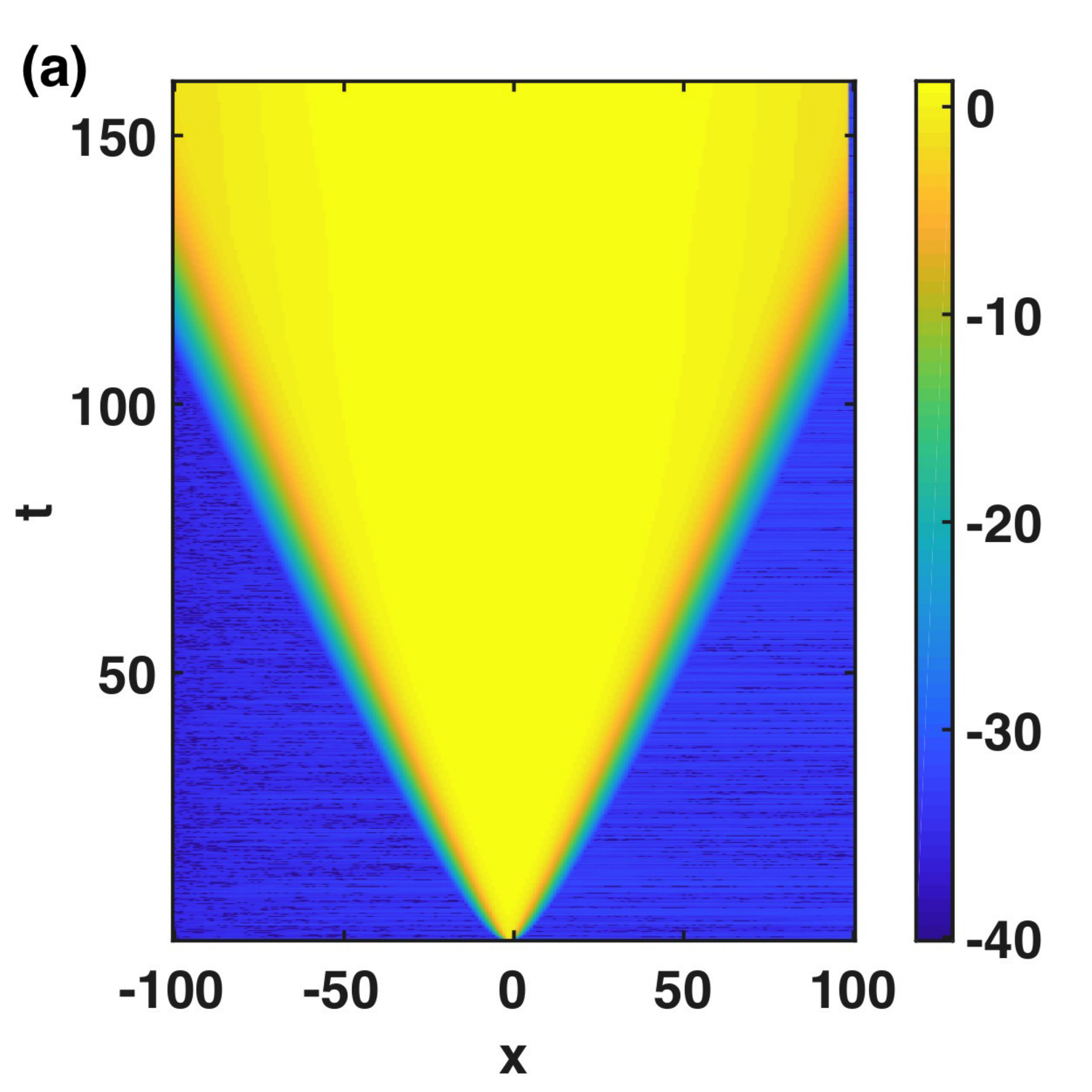}
\includegraphics[height=0.45\columnwidth, width=0.45\columnwidth]
{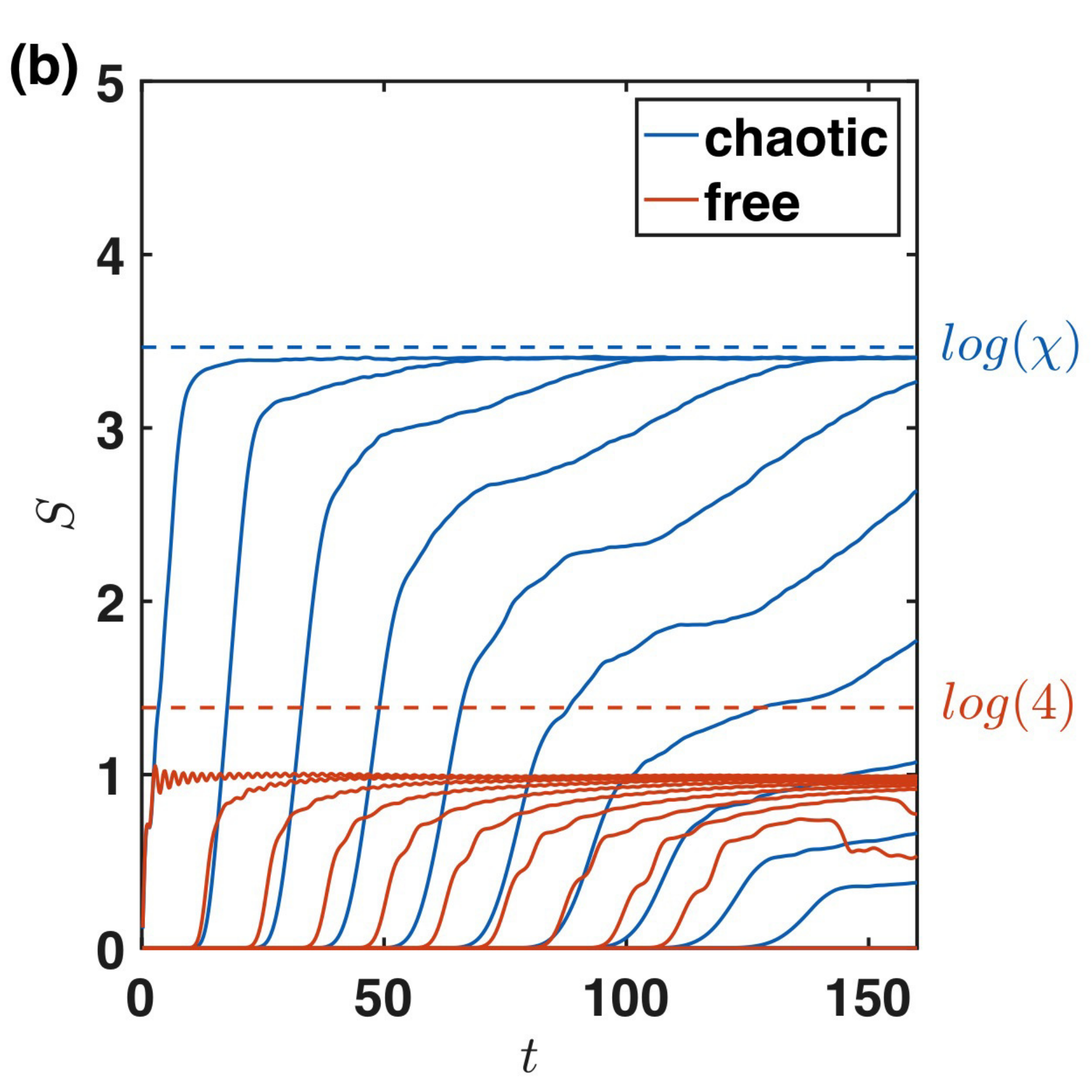}
\caption{(a) The entanglement entropy growth of a typical Heisenberg operator, say $X(t)$, is restricted in a light cone. (b) For the mixed-field Ising model, $S$ grows sharply to its maximal value $\log(\chi)$ in a short time scale $\delta t$, while for the transverse-field Ising model, $S$ saturates to a much lower value bounded by $\log(4)$. We consider spin chains with $201$ spins and set the bond dimension $\chi=32$. }
\label{fig:entropy2}
\end{figure}

We first study the operator entanglement entropy $S(x,t)$ of the Pauli operator $X_0(t)$, which is initially located at the center of spin chain. For both cases chaotic and integrable models, $S(x,t)$ exhibits the typical light cone structure shown in Fig.~\ref{fig:entropy2}(a). As demonstrated in Fig.~\ref{fig:entropy2}(b), $S(x,t)$ stays small until the wavefront arrives at time $t_0=x/v_B$ and sharply grows to its maximal $\chi$-dependent value in a rather short time-scale $\delta t \sim 1$. In the non-interacting case, $S(x,t)$ saturates to a fairly low $\chi$-independent value that is upper bounded by $\log 4$ (see Appendix~\ref{appsc:up_bound} for the proof). Therefore an MPO with bond dimension as low as $4$ is sufficient to fully represent $X(t)$ for an arbitrarily long time in the non-interacting model. On the other hand, for the chaotic case, $S(x,t)$ typically saturates to $\log(\chi)$, and the MPO cannot fully represent $X(t)$. Nevertheless, the truncation error remains inside the entanglement light cone and does not affect the early growth region.

\begin{figure}
\includegraphics[height=0.45\columnwidth, width=0.45\columnwidth]
{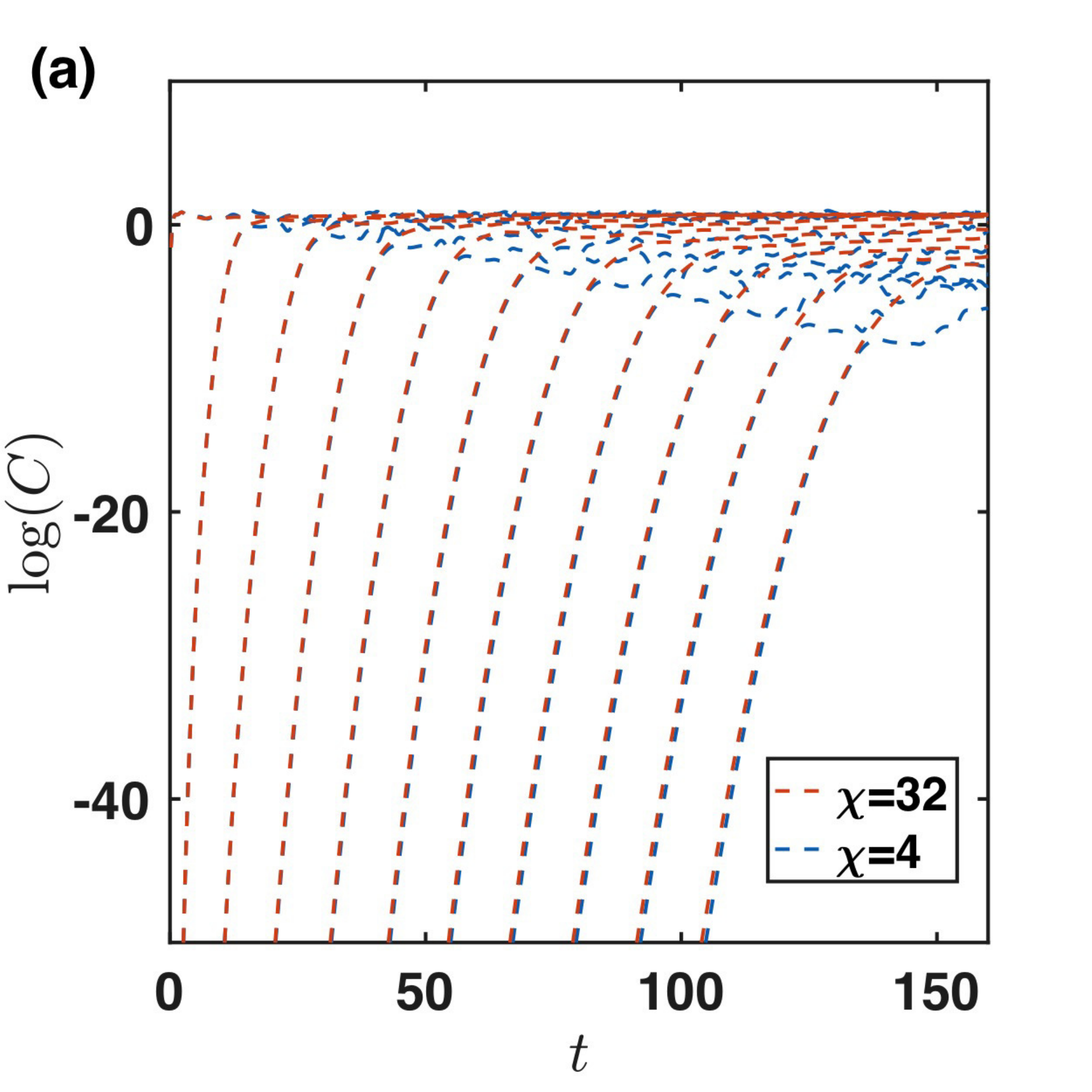}
\includegraphics[height=0.45\columnwidth, width=0.45\columnwidth]
{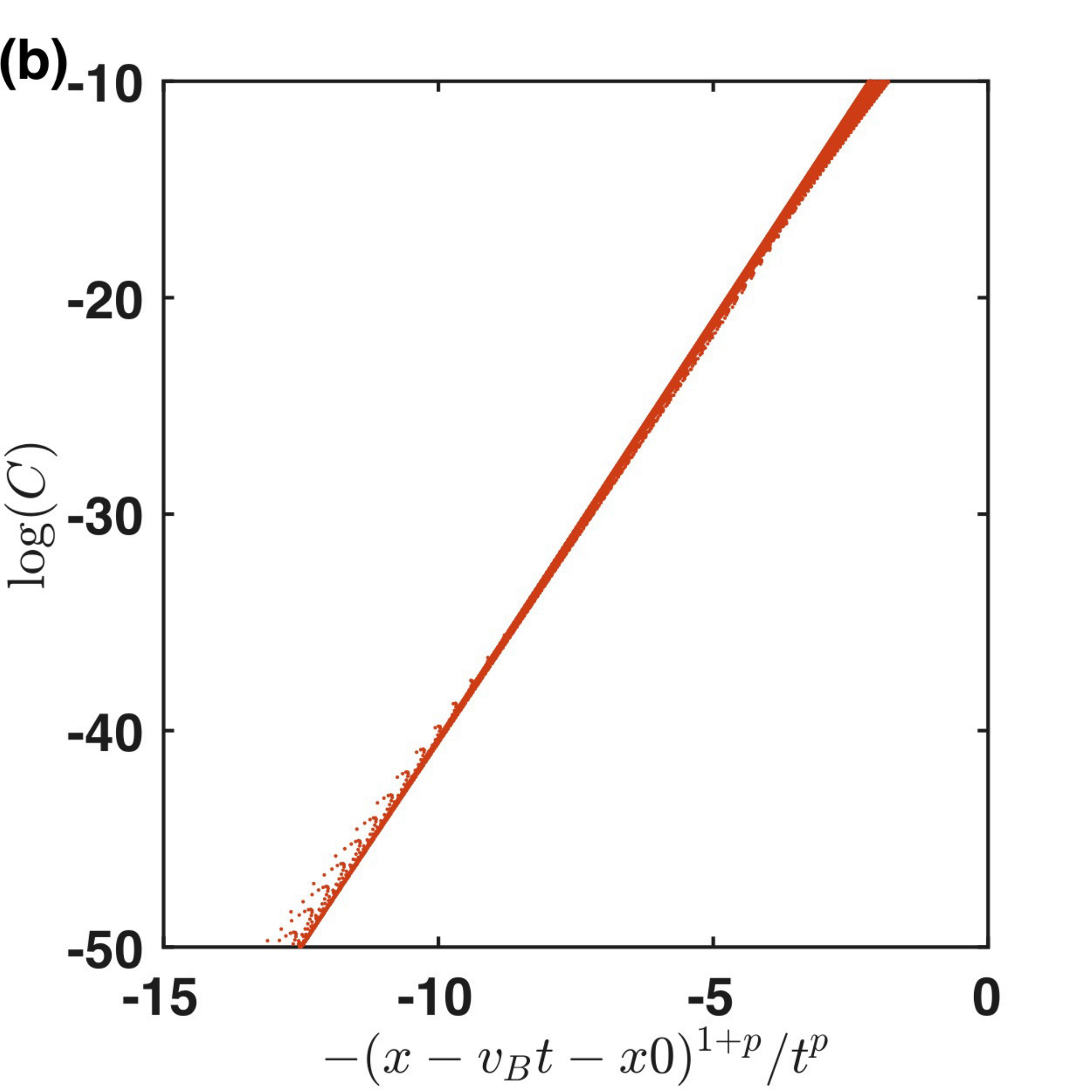}
\caption{(a) The early growth of the commutator of $X_r(t)$ and $X_{r'}$ with $r$=101 and $r'-r=0, 10, 20,\cdots,100$. An MPO with $\chi$ as small as $4$ is sufficient to capture the early growth region of $C$ from machine precision to $\sim 0.1$ across the whole chain. (b) The early growth behavior is fitted to $e^{-\lambda (x-v_Bt-x_0)^{1+p}/t^p}$ with $\lambda=3.8, p=0.67, v_B=0.67$ and $x_0=1.8$, indicating a sub-diffusively spreading wavefront.}
\label{fig:otoc_chaos}
\end{figure}

Now we use the MPO method to address the early growth region of the OTOC, the calculation of which so far has been restricted to small systems, special large $N$ models, and  weak-coupling field theories. For simplicity, we work with the infinite temperature Gibbs ensemble, so that Eq.~\eqref{eq:commutator} reads
\be
 C(r,r',t)=\frac{1}{2^L}\Tr([X_r(t), X_{r'}]^\dagger[X_r(t), X_{r'}]).
 \label{eq:cinfty}
 \ee
Fig.~\ref{fig:otoc_chaos}(a) shows the results for the mixed-field model with nonzero $h_z$ with the vertical axis on a logarithmic scale. The calculation with $\chi=4$ produces the same result as $\chi=32$ in the early growth region where $C(t)$ ranges from machine precision to $\sim 0.1$, convincingly demonstrating that MPO dynamics with a modest bond dimension $\chi$ is sufficient to capture the entire early growth region of the OTOC.
The growth rate, defined as $\frac{d}{dt} \log(C)$, slowly decreases with both the position and $\log(C)$, implying that the wavefront is spreading and no position-independent Lyapunov exponent can be directly defined. Instead, the early growth dynamics agrees very well with the form in Eq.~\eqref{eq:fitting_form} with $p\sim 0.7$ as shown by the data fitting in Fig.~\ref{fig:otoc_chaos}(b). This suggests that the wavefront spreading is sub-diffusive. We also numerically verified Eq.~\eqref{eq:fitting_form} with $p=1/2$ in the non-interacting model. Intriguingly, the growth of commutators in the isotropic spin-1/2 Heisenberg chain also approximately agrees with the form with $p\sim 0.5$ shown in Appendix~\ref{appsc:heisenberg}.

Now we discuss the behavior of the OTOC beyond the early growth region. Eq.~\eqref{eq:cinfty} can be recast as the expectation value of a local observable in terms of the operator state $\ket{X_r(t)}$
\be
C(t)=2-2\bra{X_r(t)}X_{r'}\otimes X_{r'}\ket{X_r(t)}.
\label{eq:local}
\ee
The pure state $\ket{X_r(t)}$ evolves under the 'super' Hamiltonian $\mathcal{H}=H\otimes I-I\otimes H^*$ and sits in the middle of the spectrum of $\mathcal{H}$ with approximately zero energy. The chaos of the original model $H$ directly translates into chaos for $\mathcal{H}$, and thus thermalization implies that the second term approaches $\Tr (X_{r'}\otimes X_{r'}) =0 $ as $t\rightarrow \infty$. Therefore $C(t)$ is expected to saturate to $2$ in the late time limit for chaotic models.

\begin{figure}
\includegraphics[height=0.45\columnwidth, width=0.45\columnwidth]
{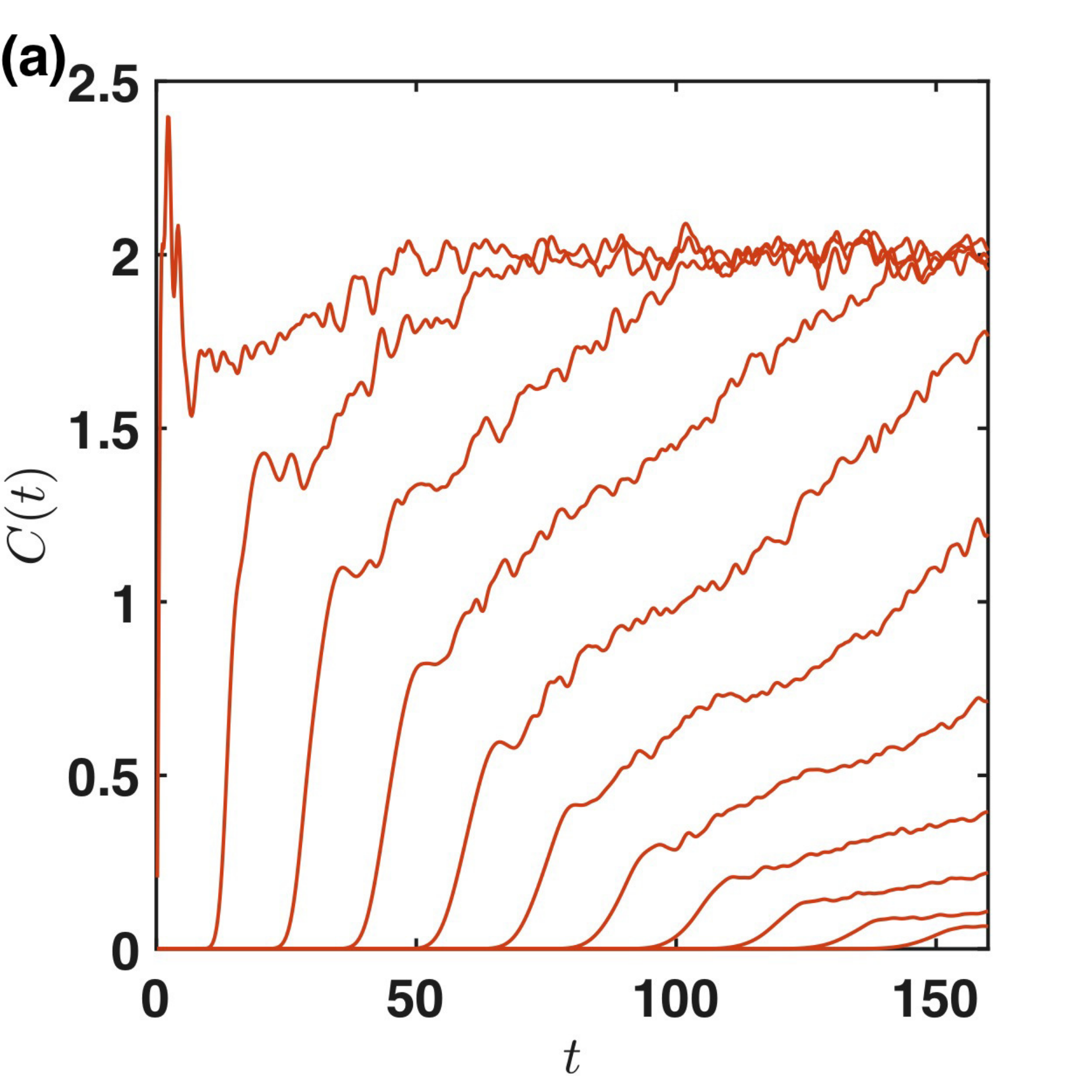}
\includegraphics[height=0.45\columnwidth, width=0.45\columnwidth]
{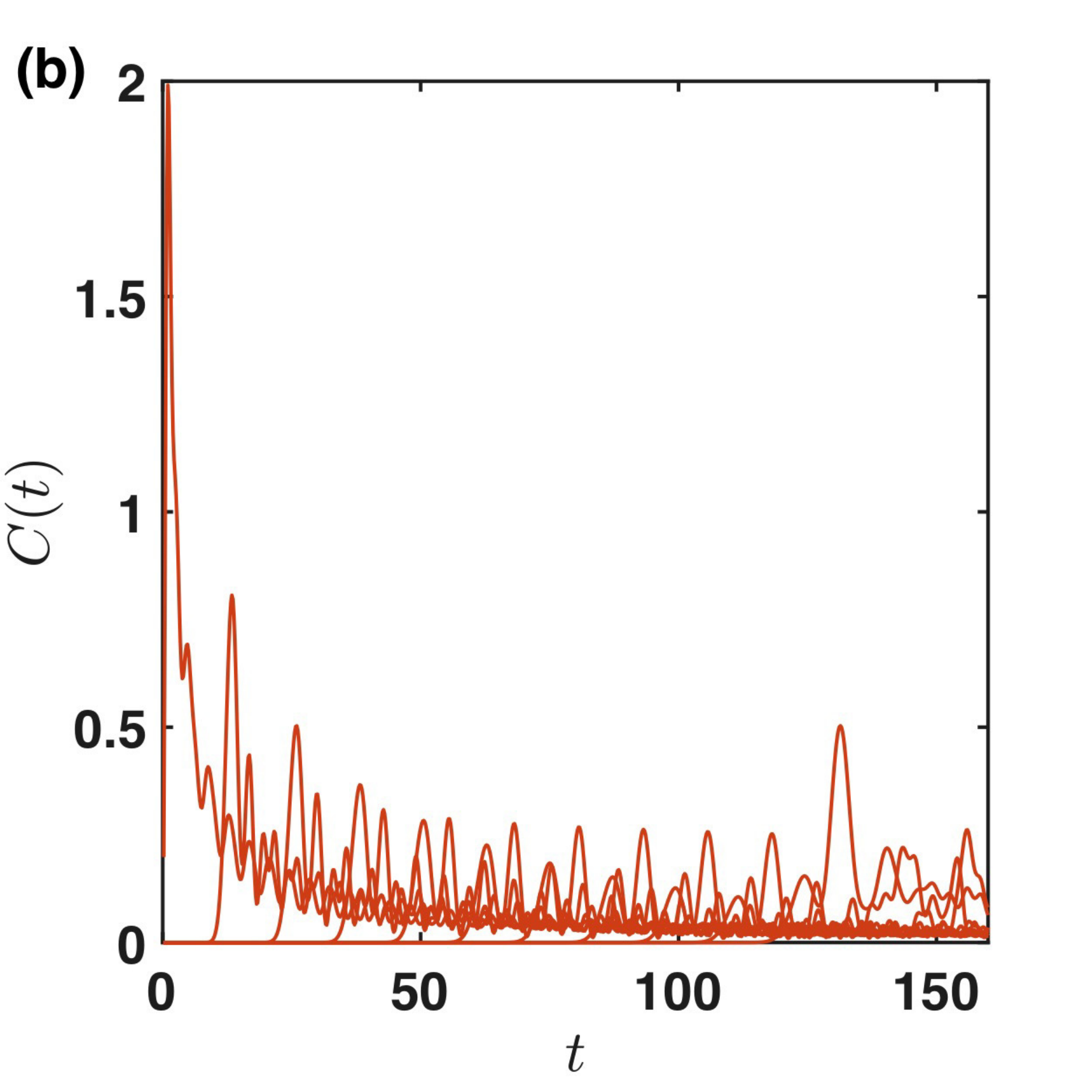}
\caption{(a) The OTOC for the mixed-field model. The commutator exhibits two growth regions. In the second region, the growth is much slower and relaxes to the final value of $2$. The MPO result thus qualitatively matches expectations even inside the light cone. (b) In the case of the transverse-field model, an MPO with $\chi=4$ is sufficient to obtain $C(t)$ for arbitrarily long time because of the bounded entanglement. The commutator decays to $0$ in the late-time limit, showing no sign of scrambling as expected.}
\label{fig:c_int}
\end{figure}

Even though our MPO approach cannot precisely capture the operator $X_r(t)$ in the long-time limit within the light cone because of the saturation of the entanglement entropy, it still qualitatively produces the correct late time behavior of $C(t)$, as shown in Fig.~\ref{fig:c_int}(a). Furthermore, and quite surprisingly, the figure clearly shows that the growth of the OTOC separates into two different time regions: the early growth region and a second growth region with a much slower growth rate. Two recent works~\cite{Khemani2017, Rakovszky2017} on local random circuit models with an onsite symmetry argued that one get a second growth region (in fact, a slow decay to the late time limit) as a consequence of conservation laws. In our case, only energy is conserved and we observe that our secondary growth region occurs even away from late time saturation of $C(t)$. On the other hand, the transverse-field Ising model, due to the absence of thermalization,
exhibits completely different behavior in the late-time regime, decaying to zero with a universal $1/t$ tail as derived for generic non-interacting systems in Appendix~\ref{appsc:noninteracting}.

\textit{Discussion---}In this work, we presented both a theoretical conjecture and a low-cost numerical method that can address the early growth of scrambling in generic local quantum many-body systems. The method is based on the tight connection between operator growth and scrambling, and it opens up numerous directions for future research. Regarding quantum chaos, a crucial question is whether different values of $p$ define different universality classes of scrambling. Our analysis shows that integrable models, various chaotic energy conserving systems, and random circuit models can in fact distinguished by the value of $p$ associated to the early growth of commutators. One may further wonder if $p$ is related to the level statistics of the energy spectrum or to other physical properties of the system. We also observed a small upward drift of the extracted $p$ value when including larger distance data, so one may need to explore even larger size systems to determine the true asymptotic structure.

Among other interesting directions, we are currently applying the MPO method to study scrambling study scrambling at finite temperature and in systems with disorder. In both cases, we suspect that the universal form Eq.~\eqref{eq:fitting_form} should be modified. At finite temperature, the bound on chaos \cite{Maldacena2016} should enter in an interesting way. For many-body localized systems exhibiting slow scrambling, the ballistic wavefront assumption underlying Eq.~\eqref{eq:fitting_form} is no longer valid, but the slow scrambling can still be captured with MPO techniques.

Still another interesting direction is to view the time evolution of the tensors defining the MPO as an auxiliary classical dynamical system associated to the quantum growth of operators. In fact, there is a family of auxiliary dynamical systems indexed by bond dimension, and our analysis shows that this family better and better approximates the quantum chaotic growth of operators. Perhaps the Lyapunov exponents of this associated classical dynamics allow us to define a spectrum of Lyapunov exponents for the associated quantum system. We are currently studying the time-dependent variational principle version of MPO dynamics, which has an associated symplectic structure, to address this possibility.

Furthermore, our MPO technique is not restricted to Hamiltonian systems but can also be applied to driven systems, since the bound in Eq.~\eqref{eq:bound} still holds. In a preliminary study of a Floquet version of the mixed-field Ising model, we observed two wavefronts in the early growth region of the squared commutator. The first chaos wave appears similar to that in the Hamiltonian system, except that it decays and gives way to a second chaos wave with a smaller $v_B$. It is interesting to extract and carefully study the second wavefront in the Floquet model; perhaps the decay of the first chaos wave is associated with the loss of energy conservation. This is challenging because the second chaos wave lies within the light cone of the first wave and hence is more sensitive to truncation errors.

From an experimental perspective, Eq.~\eqref{eq:local} allows one to interpret the OTOC as a local observable in a kind of quench experiment. To measure it, one must first engineer the thermofield double state $\Phi$ (which is just Bell pairs) on, say, a ladder of spins. Then the state is quenched by acting with $X_r\otimes I$ on the center rung of the ladder and evolving under $\mathcal{H}$. Finally, the OTOC can be read off as the expectation value of $X_{r'}\otimes X_{r'}$. This scheme based on operator dynamics has two advantages. First, it does not require evolving the state back and forth in time, although the required super Hamiltonian $\mathcal{H}$ must still be engineered. Second, this scheme is not biased toward any initial state since it directly measures OTOC in the infinite temperature Gibbs ensemble. This scheme is a complement to the numerical algorithm presented here, since together they may plausibly give access to several different dynamical regimes of the OTOC.

\textit{Acknowledgement-} We thank Jim Garrison, Dan Roberts, Alexei Kitaev, and Douglas Stanford for interesting discussions. This material is based upon work supported by the Simons Foundation via the It From Qubit Collaboration, by the Air Force Office of Scientific Research under award number FA9550-17-1-0180, and by the NSF Physics Frontier Center at the Joint Quantum Institute (PHY-1430094).

\bibliography{otoc}

\appendix

\section{Discussion of different possible OTOC growth forms}
\label{appsc:growth_form}

Throughout this appendix we focus on the case of OTOCs of two local operators, for simplicity assumed to be Hermitian and unitary (like a local Pauli operator), with time evolution generated by a geometrically local Hamiltonian. Operator $W$ is placed at position zero while operator $V$ is at position $x$. The OTOC is
\be
F(x,t) = \langle W_0(t) V_x W_0(t) V_x \rangle,
\ee
and the squared commutator is
\be
C(x,t) = \langle [W_0(t),V_x]^\dagger [W_0(t), V_x]\rangle = 2(1 -  \Re[F(x,t)]).
\ee

Focusing on the early growth region, where $C\ll 1$, many possible forms for $C$ have been proposed. We first catalog a number of the different forms:
\begin{enumerate}
  \item{Perturbative:} $C(x,t) \sim \frac{(at)^{bx}}{(bx)!}$
  \item{Exponential/Holographic:} $C(x,t) \sim \exp\left(\lambda t - \mu x\right)$
  \item{Random circuit:} $C(x,t) \sim \exp\left(-\frac{(x-vt)^2}{D t} \right)$
  \item{Growth/Diffusion:} $C(x,t) \sim \exp\left(\lambda t -\frac{x^2}{D t}\right)$
  \item{X/S:} $C(x,t) \sim \exp\left(- \lambda \frac{(x-vt)^{1+p}}{ t^p} \right)$
\end{enumerate}
This list is not necessarily exhaustive, but it does include a representative sample. Depending on the physical system, different forms may apply in different dynamical regimes.

The perturbative form arises from literal perturbation theory. For a spin chain with a local Hamiltonian, the coefficient of the first non-vanishing term in the commutator $[W_0(t),V_x]$ is proportional to $t^x/x!$, i.e., to a high power of $t$ arising from the high order in perturbation theory needed to find a non-vanishing contribution to the commutator. As implied by this discussion, such a form might be expected to approximately hold at very early time in almost any spin chain.

The exponential/holographic form is obtained from black hole computations in AdS/CFT and in coupled SYK clusters. It can also be obtained near the wavefront in a saddle point analysis. As discussed \cite{Chowdhury2017,Patel2017, Gu2017}, in a variety of situations one can approximate the spatial dependence of the commutator as the Fourier transform of an exponentially growing function of time where the growth rate is momentum dependent. In one dimension this reads
\be
C(x,t) \sim \int dp \,\tilde{C}(p) e^{\lambda(p) t + i p x}.
\ee
This integral may be analyzed by attempting to deform the integration contour to the saddle point $p=p_s$ obtained from
\be
\frac{\partial \lambda}{\partial p}\bigg|_{p=p_s} t + i x =0
\ee
where we note that $p_s$ depends only on $x/t$.

When the coefficient $\tilde{C}(p)$ is non-singular, the contour may be deformed all the way to the saddle point. Then the integral is approximated by
\be
C(x,t) \sim \tilde{C}(p_s) e^{\lambda(p_s) t + i p_s x}.
\ee
The wavefront velocity is defined by $v_B = x/t$ and is given by
\be
\lambda(p_s) + i p_s v_B = 0.
\ee
Defining $t = x/v_B + \delta$ and expanding in small $\delta$, the commutator becomes
\be
C(x,t) \sim \tilde{C}(p_s) e^{\lambda(p_s) \delta + O(\delta^2/t)}
\ee
which is of the exponential form.

When the coefficient $\tilde{C}(p)$ is singular, say with a pole at $p=p_0$, then at large $x/t$ one typically finds that $\Im(p_0) < \Im(p_s)$. In this case, as the countour is deformed to the saddle point, we also pick up a pole contribution. Since it has smaller imaginary part, it dominates the integral giving
\be
C(x,t) \sim \left(\text{Res}_{p=p_0} \tilde{C}(p) \right) e^{\lambda(p_0) t + i p_0 x}.
\ee
In the simplest case where $p_0$ is pure imaginary, one again recovers the exponential form. This situation obtains in both holographic models and coupled SYK clusters.

The random circuit form is obtained from an analysis of local random circuit models, in which the dynamics is defined by alternating even and odd layers of Haar random two spin unitaries. One interesting feature of the early time growth region is that, while a wavefront exists, the size of the wavefront spreads diffusively. Recently, the addition of onsite symmetries to the random circuit models was considered. One significant modification that arises is a slow late time decay to $C\sim 2$ due to a slow hydrodynamic mode.

The growth/diffusion form arises in several analyses, most prominently in the study of disordered systems. It can also be viewed as an approximation to the situation in which we have a momentum dependent growth exponent. If $\lambda(p)$ is approximated by $\lambda_0 - \lambda_2 p^2$, then the growth/diffusion form is obtained. This form may not be valid at very large $x/t$ since the saddle point momentum increases like $p_s \sim i (x/t)$ which may move outside the regime of validity of the quadratic approximation to $\lambda(p)$. In this case, we return to the general saddle point analysis above.

One important constraint all these forms must obey is microscopic causality, either due to Lorentz invariance or, say, a Lieb-Robinson bound. Furthermore, at finite temperature in an energy conserving system, we expect the growth rate of the commutator to be bounded by
\be
\frac{1}{C} \frac{d C}{dt} \leq 2\pi T,
\ee
which is the chaos bound. Note that there is a small assumption in this statement, since the chaos bound technically applies to the thermally regulated OTOC, but we believe the same bound applies to the non-regulated commutator with one small caveat. When the chaos bound is saturated, the non-thermally regulated OTOC has a leading part which is pure imaginary, so the commutator, which involves $\Re(F)$, is not sensitive to the leading growth. In this case, $C$ may grow at rate $2 \times 2\pi T$. A similar feature occurs for some systems in the semi-classical limit.

The X/S form proposed here is able to encompass many of the other forms, as well as the growth of $C$ in non-interacting particle models (see Appendix~\ref{appsc:noninteracting}). It can be motivated by considering an early growth wavefront region defined by $x,t$ large with $x/t$ fixed and larger than $v_B$, the wavefront velocity. It is convenient to analyze $C$ using the variables $x-v_B t$ and $t$. Then it is natural to guess that $C$ grows quasi-exponentially, with a form
\be \label{eq:XSform_general}
\log C = - \gamma \frac{(x-v_B t)^\alpha}{t^\beta} + \cdots.
\ee
Requiring that $\log C$ have a non-infinite and non-zero time derivative at fixed $x/t$ requires $\alpha = \beta+1$. This same constraint on $\alpha$ and $\beta$ can be obtained more carefully as follows.

Fixing $ x/t = v  > v_{LR} \geq v_B$, where $v_{LR}$ is the Lieb-Robinson velocity, the Lieb-Robinson bound, $C \leq C_0 e^{\kappa (t - x/v_{LR})}$, gives
\be
C \leq C_0 e^{- \kappa \frac{v-v_{LR}}{v v_{LR}} x},
\ee
meaning $C$ must decay at least exponentially fast. Plugging the same limit into Eq.~\eqref{eq:XSform_general} gives
\be
\log C = - \gamma x^{\alpha -\beta} \frac{(v -v_B)^\alpha}{v^\beta}.
\ee
Requiring at least exponential decay gives $\alpha \geq \beta + 1$.

On the other hand, requiring that $\frac{d}{dt} \log C$ be finite at fixed $x/t$ with $x,t$ large gives an upper bound on $\alpha$. The time derivative is
\be
\frac{d}{dt} \log C = \gamma \alpha v_B \frac{(x-v_B t)^{\alpha-1}}{t^\beta} + \gamma \beta \frac{(x-v_B t)^\alpha}{t^{\beta+1}}.
\ee
To have $\frac{d}{dt} \log C <\infty $ as $x,t\rightarrow \infty$ thus requires $\alpha \leq \beta +1$. Note that the finiteness of the logarithmic derivative is motivated by the chaos bound, although we are arguing for it even for integrable systems, for energy non-conserving systems, and for infinite temperature. We would generally expect that $\frac{d}{dt} \log C < J$, where $J$ is some microscopic possibly time-dependent coupling in the Hamiltonian.

Since we have both $\alpha \geq \beta +1$ and $\alpha \leq \beta +1$, it follows that $\alpha=\beta+1$ in which case Eq.~\eqref{eq:XSform_general} is the X/S form.

\section{Operator entanglement, OTOCs, thermofield doubles, and holography}
\label{appsc:operator_entanglement}

In this appendix, we relate the operator entanglement of the Heisenberg operator $W(t)$ to OTOCs involving $W(t)$. For concreteness, we consider a spin chain of $n$ spins with open boundary conditions and we take $W$ to be a unitary operator supported on a region $R_W$. Divide the spin chain into two regions, region $A$, which contains $R_W$, and region $B$, which is the complement of $A$. The Heisenberg operator is $W(t) = U^\dagger W U$ where $U= e^{- i H t}$.

To define the operator entanglement of $W(t)$, consider two copies of the spin chain in which each spin ($r$) is maximally entangled with its partner spin ($\tilde{r}$):
\be
|\Phi \rangle = \prod_{r=1}^n \frac{|00\rangle_{r \tilde{r}}+|11\rangle_{r\tilde{r}}}{\sqrt{2}}.
\ee
Now define the operator state
\be
|W(t)\rangle = (W(t) \otimes I ) |\Phi \rangle.
\ee
The operator entanglement of $W(t)$ for region $A$ is then the entanglement of region $A$ and its partner $\tilde{A}$ in the state $|W(t)\rangle$, $S(A\tilde{A},|W(t)\rangle)$.

At time zero, when $W(t)=W$, the operator entanglement of $W$ is zero provided $R_W \subset A$. Similarly, if $O_B$ is a Pauli string supported in region $B$, then the OTOC, defined as $F(t)=\text{Tr}(\rho_\infty W(t)^\dagger O_B W(t) O_B )$, also obeys $F(t=0)=1$. We want to bound the operator entanglement of $W(t)$ in terms of how much $F(t)$ differs from one.

The following useful operator inequality will be used multiple times in the derivation,
\be
|\Tr(\rho_\infty A^\dagger U_1 A U_2)|\leq\Tr(\rho_\infty A^\dagger A)
\label{eq:inequality}
\ee
where $U_1$ and $U_2$ are unitary. This inequality comes after the fact that the expectation value of unitary operator is always smaller than 1.

To proceed, we first adapt the result derived in a recent work \cite{Hosur2016} to relate the operator entanglement of $W(t)$ to an $8$-point OTOC. The main idea here is to replace the time evolution $U$ with a new time evolution given by $W(t)=U^\dagger W U$. Ref.~\cite{Hosur2016} related the operator entanglement of $U$ to the average over operators of a $4$-point OTOC. Here we find that
\be
e^{-S^{(2)}(A\tilde{A})}= \frac{1}{4^n}\sum_{O_A,O_B} F_8
\ee
where
\be
F_8 =  \text{Tr}\left(\rho_\infty O_A W(t)^\dagger O_B W(t) O_A W(t)^\dagger O_B W(t)\right),
\ee
$S^{(2)}(A\tilde{A})$ is the second R\'enyi entropy of $A\tilde{A}$, $O_A$ and $O_B$ are Pauli strings in $A$ and $B$, respectively, and we have used the fact that $|A|+|B|=n$.

Now we need to relate $F_8$, which depends on $O_A$ and $O_B$, to a $4$-point OTOC denoted $F_4$,
\be
F_4 (O_B)= \text{Tr}\left(\rho_\infty O_B W(t)^\dagger O_B W(t)\right),
\ee
which depends only on $O_B$. It is crucial to remove the dependence on $O_A$ since only $O_B$ approximately commutes with $W(t)$. We find
\be
F_8(O_A, O_B)\geq 4 F_4(O_B)-3
\ee

This is proven by rewriting $O_A W(t)^\dagger O_B W(t)$ as
\be
O_A W(t)^\dagger O_B W(t) = O_A O_B + O_A W(t)^\dagger [O_B , W(t)].
\ee
Inserting this expression into $F_8$ and using $[O_A,O_B]=0$ and $O_A^2=1$ gives
\be
F_8=2F_4-1+\text{Tr}\left(\rho_\infty O_A  [ O_B ,W(t)] O_A [ O_B ,W(t)]\right).
\label{eq:f4}
\ee
Using the inequality Eq.~\eqref{eq:inequality}, the second term in this expression obeys
\bea
&\text{Tr}\left(\rho_\infty O_A  [ O_B ,W(t)] O_A [ O_B ,W(t)]\right) \\
& \ \ \ \ \geq \text{Tr}\left(\rho_\infty [O_B,W(t)] [O_B,W(t)] \right),
\eea
but the right hand side is nothing other than $2F_4-2$, and we arrive at Eq.~\eqref{eq:f4}.

Here $F_4$ contains Pauli strings in region $B$, and we want to relate it to OTOCs depending on only local operators. To do so, we rewrite $F_4$ as,
\bea
F_4=& \Tr(\rho_\infty W(t) O_B' W(t) O_B') \\
        &+\frac{1}{2}\Tr(\rho_\infty O_B' [W(t), O_i]O_B'[W(t), O_i]),
\eea
where the local operator $O_i$ acts on the boundary of $B$, and $O_B'$ acts on the rest, denoted as $B'$, so $O_B=O_iO_B'$ and $[O_i, O_B]=0$.  From the inequality Eq.~\eqref{eq:inequality}, the second term is larger than $\frac{1}{2}\Tr(\rho_\infty  [W(t), O_i][W(t), O_i])$. The we get
\bea
F_4(O_B)\geq &\Tr (\rho_\infty W(t)O_{B'} W(t) O_{B'})\\
&-\frac{1}{2}\Tr (\rho_\infty [W(t),O_{i}]^\dagger [W(t), O_{i}]).
\eea
We perform the trick repeatedly and eventually obtain,
\be
F_4(O_B)\geq 1-\frac{1}{2}\sum\limits_{i\in B} ||[W(t), O_i]||^2,
\ee
where $O_B=\prod \limits_{i\in B}O_i$, and $O_i$, commuting with each other, acts on each site of region B. Therefore, $F_8$ can be bounded by the squared commutator between $W(t)$ and $O_i$ as
\be
F_8(O_A, O_B)\geq 1-2\sum\limits_{i\in B} ||[W(t), O_i]||^2
\ee

By performing the operator average of $O_A$ and $O_B$, the bound on $F_8$ (valid for each term in the average) gives
\be
e^{-S^{(2)}(B\tilde{B})}  \geq 1-\frac{1}{2}\sum\limits_{i\in B, O_i} ||[W(t), O_i]||^2.
\ee
 For $\sum\limits_{i\in B, O_i} ||[W(t), O_i]||^2<2$,  we have
\be
S^{(2)}(B\tilde{B}) \leq -\log\left( 1-\frac{1}{2}\sum\limits_{i\in B, O_i} ||[W(t), O_i]||^2\right)
\ee
This is rigorous independent of how much region $B$ and the support of $W(t)$ overlap.
If we restrict region $B$ far away from $R_W$, then each term in the summation is small and decays exponentially or faster with the shortest distance between $B$ and $R_W$, according to the Lieb-Robinson bound. In this case, we can expand the logarithm and bound the operator entanglement by the squared commutator of $W(t)$ with local operators acting on the entanglement cut as
\be
S^{(2)}(B\tilde{B})\leq \frac{1}{2(1-e^{-2a})}\sum\limits_{O_i} ||[W(t), O_i]||^2
\ee
where $a$ is the spatial exponential decay constant.

Once the wavefront has past the entanglement cut, we expect the operator entanglement to grow linearly with time. Given an MPO with fixed bond dimension $\chi$, the physics can only be accurately captured for a finite temporal extent $\delta t$ inside the wavefront. The value of $\delta t$ may be estimated via $S(x,t=x/v_B + \delta t) \sim \log \chi$. Note also that truncation errors propagating inside the light cone can also be problematic even if the entanglement is still within the constraints set by the bond dimension.

Now the state $|\Phi\rangle$ used above is a purification of the infinite temperature state, i.e., the maximally mixed state, of the spin chain. One can generalize the above calculation to the situation where $|\Phi\rangle$ is replaced by the thermofield double state at temperature $T$,
\be
|T\rangle= \sum_{\alpha} \sqrt{\frac{e^{-E_\alpha/T}}{Z}} |E_\alpha \rangle_{AB} \otimes |E_\alpha \rangle_{\tilde{A}\tilde{B}}.
\ee
One then defines a new operator state
\be
|W(t),T\rangle=(W(t)\otimes I)|T\rangle,
\ee
where $|W(t),T=\infty\rangle=|W(t)\rangle$. Finally, one relates the increase in the entanglement of $A\tilde{A}$ in state $|W(t),T\rangle$ relative to state $|W(0),T\rangle$ to OTOCs at temperature $T$ involving $W(t)$ and $O_B$.

The operator entanglement in the non-interacting limit is analyzed in Appendix~\ref{appsc:up_bound}. To gain intuition for how the operator entanglement behaves with interactions, let us consider the example of a strongly coupled gauge theory with an AdS/CFT dual consisting of Einstein gravity. We focus on the case of CFT$_{1+1}$/AdS$_{2+1}$ to match with our spin chain setting, but the results can be generalized to higher dimensions.

The AdS/CFT dual of the state $|W(t),T\rangle$ is a so-called shockwave geometry (see e.g.,~\cite{Shenker2014}). The starting point is the unperturbed BTZ black hole metric in asymptotically AdS$_{2+1}$,
\be
ds^2 = - \frac{r^2 - R^2}{\ell^2} dt^2 + \frac{\ell^2}{r^2-R^2} dr^2 + r^2 d\phi^2,
\ee
where $\ell$ is the AdS radius, $\phi \sim \phi+ 2\pi $ is an angular coordinate, $R^2 = 8 G_N M \ell^2$ is the black hole radius ($M$ is the mass), and the inverse temperature is $\beta = \frac{2\pi \ell^2}{R}$. This metric describes an entangled state between two CFTs, the two sides of the wormhole, where each CFT is defined on a circle and is in a thermal state of temperature $T = 1/\beta$.

The shockwave metric is more conveniently described using Kruskal coordinates $u,v$. The metric reads
\be
ds^2 = \frac{-4 \ell^2 du dv + R^2(1-uv)^2 d\phi^2}{(1+uv)^2}.
\ee
The coordinates are related by
\be
\frac{v+u}{1+uv} = \frac{\sqrt{r^2 - R^2}}{R} \sinh \frac{R t}{\ell^2}
\ee
and
\be
\frac{v-u}{1+uv} = \frac{\sqrt{r^2 - R^2}}{R} \cosh \frac{R t}{\ell^2}.
\ee

The shockwave is applied by evolving backward in time to $t=-t_w$, applying operator $W$, then evolving forward in time to $t=0$. This corresponds to applying the Heisenberg operator $W(-t_w)$. We first review the case when $W$ is non-local, then we discuss the case where $W$ is a local perturbation. This means we will first ignore the spatial dependence (which is only really interesting in a system, like systems with a classical gravity dual, in which there are many local degrees of freedom) and then later consider the spatial structure.

The bulk description of the perturbation is the addition of a small amount of energy $\delta E$ in the distant past. As this energy falls towards the black hole, it is blue shifted, and near the horizon the proper energy is
\be
E_{\text{hor}} \sim \delta E\frac{\ell}{R} e^{R t_w/\ell^2} = \delta E\frac{\ell}{R} e^{2\pi t_w/\beta}.
\ee
Even when $\delta E \sim T$ is a small thermal energy scale, the proper energy near the horizon can be large if $t_w$ is large.

It is useful to consider the thermodynamic limit in $\delta E/M$ is small and $t_w$ is large with
\be
\alpha = \frac{\delta E}{4 M} e^{2\pi t_w/\beta}
\ee
fixed. In this limit, the shockwave geometry is obtained by matching two different black hole geometries of equal mass $M$ and Kruskal coordinates $u,v$ and $\tilde{u},\tilde{v}$. The matching is done across the infalling energy shell which hugs the horizon; the condition is $\tilde{v} = v+ \alpha$ and $\tilde{u}=u$. The shockwave metric may be written
\be
ds^2 = \frac{-4 \ell^2 du dv + R^2(1-u(v+\alpha \theta(u)))^2 d\phi^2}{(1+u(v+\alpha \theta(u)))^2}.
\ee

The analog of the OTOC may be obtained from Eq.~\eqref{eq:local} by considering the correlator of two local operators, one in each copy of the system (one on each side of the wormhole). Suppose the CFT operators $V_L$ and $V_R$ are dual to bulk fields with mass $m$ (corresponding primary operators in the CFT of dimension $\Delta \sim m\ell$). Then using a geodesic approximation, the correlator is
\be
\langle W(-t_w),T| V_L \otimes V_R | W(-t_w),T\rangle \sim \left(\frac{1}{1+\alpha/2}\right)^{2 \Delta},
\ee
where we have normalized by the value at $\alpha=0$. Expanding in small $\alpha$ gives
\be
\text{OTOC} \sim 1 - \Delta \alpha + \cdots.
\ee

What about the operator entanglement? Let $A$ and $\tilde{A}$ denote corresponding regions on the two sides of the wormhole and let $B$ and $\tilde{B}$ be their complements on each side. The entanglement can be computed from the Ryu-Takayanagi formula. In the thermodynamic limit in which the size of the CFT circle in units of the thermal length $\beta$ is large, the entanglement is
\be
S(A \tilde{A}) =S(B\tilde{B}) = \frac{\ell}{G_N}\left( \log \frac{2r_c}{R} + \log (1+\alpha/2) \right),
\ee
where $r_c$ is a large distance cutoff in the bulk that is dual to a short distance cutoff of $\ell^2/r_c$ in the boundary CFT. The central charge of the CFT is $c =\frac{3\ell}{2 G_N}$. We see that the difference between the $\alpha=0$ operator entanglement and the $\alpha \neq 0$ operator entanglement is
\be
\Delta S(A\tilde{A}) = \frac{\ell}{G_N} \log (1+\alpha/2)
\ee
which is approximately $\Delta S \sim c \alpha $ for small $\alpha$. Hence, when $1- \text{OTOC}$ is small, the operator entanglement is also small.

Now we turn to the case of a localized perturbation. In this case, we can still use the shockwave metric in Kruskal coordinates, but the shift $\alpha$ now depends on space, $\alpha = \alpha(\phi)$. Plugging the shockwave metric into Einstein's equation and taking the thermodynamic limit of a large black hole, $R/\ell \gg 1$, one finds that $\alpha(\phi)$ obeys
\be
\left( \frac{\partial^2}{\partial \phi^2} - \frac{R^2}{\ell^2} \right) \alpha = \alpha_0 \delta(\phi),
\ee
where $\alpha_0 \propto (\delta E/4M) e^{2\pi t_w/\beta}$ and we have taken $W$ to act at $\phi=0$. Since $R/\ell$ is large, the solution is approximately
\be
\alpha(\phi) \sim \alpha_0 e^{ - R\phi/\ell}
\ee
for $\phi >0$. Identifying the effective position coordinate as $x =\ell \phi$, we see that the exponential decay of $\alpha(\phi)$ with $\phi$ is balanced by the exponential growth of $\alpha_0$ with $t_w$ when $ x=t_w$, indicating ballistic growth of $W(-t_w)$ with $v_B=1$ (the speed of light).

There is an important subtlety here. Since $v_B=1$ in this (1+1)-dimensional Lorentz invariant theory, microscopic causality requires the commutator to exactly vanish for $x > v_B t$. However, since we are studying the operator entanglement at finite temperature, the correlator $\langle W(-t_w),T| V_L \otimes V_R | W(-t_w),T\rangle$ is related to a thermally regulated OTOC which does not obey an exact causality bound. In such a case, we would use the MPO/MPS methods to describe $|W(-t_w),T\rangle$ instead of the non-thermally regulator commutator since the latter has no early growth region (being identically zero for $x>v_B t$).

Since the thermally regulated OTOC $\langle W(-t_w),T| V_L \otimes V_R | W(-t_w),T\rangle$ still obeys $1-\text{OTOC}\sim \alpha(\phi)$ for small $\alpha(\phi)$, we must simply show that the operator entanglement is similarly small. In fact, we do not need to calculate the minimal area (in this case, length) surface as required by the Ryu-Takayanagi formula, since we can bound its area (length) by a term proportional to $\alpha(\phi)$. Indeed, taking the same constant $\phi$ curve that was the minimal surface for uniform $\alpha$ (but which is no longer necessarily minimal) shows that
\be
\Delta S \leq \frac{\ell}{G_N} \log(1 + \alpha(\phi)/2).
\ee
Expanding in small $\alpha(\phi)$ shows that the operator entanglement is small when $1-\text{OTOC}$ is small, even in the case of a local perturbation $W$. In fact, the factors of $G_N$ actually cancel so that the change in the operator entropy is order unity when $t < x/v_B$.

This analysis was performed in a (1+1)-dimensional holographic CFT (meaning a CFT with classical Einstein gravity dual), but it can easily be generalized to holographic CFTs in any dimension and to other kinds of holographic systems. The general conclusion is the same: While the operator entanglement has a non-zero value at non-infinite temperature due to the non-zero thermal correlation length, the growth of the operator entanglement with time always tracks the thermally regulated OTOC.

\section{Early growth of the scrambling and wavefront spreading in non-interacting systems}
\label{appsc:noninteracting}
In this appendix, we demonstrate that the square commutator has the following form for generic non-interacting system with translational symmetry,
\be
C(x,t)\sim e^{-\lambda \frac{(x-v_Bt)^{3/2}}{t^{1/2} }}
\label{eq:free}
\ee

In non-interacting system, the dynamics of the single-particle creation/annihilation is simply $c(k,t)=e^{i \varepsilon(k)t} c(k,t)$. In the real space, we have $c_x(t)=\sum\limits_y u(x-y, t) c_y$. The function $u(x,t)$ which is the  fourier transformation of $e^{i \varepsilon(k) t}$, directly enters the commutator and controls its growth. For example, the squared commutator of two local density operators is $|u(x-y,t)|^2-|u(x-y,t)|^4$.  Thus it is interesting to analyze the function $u(x)$, which surprisingly has some universal features independent of the dispersion.

We perform the standard saddle point analysis and expand the function around $k_0$ as
\bea
u(x,t)&=e^{i (\varepsilon(k_0)t-k_0 x)}\\
&\int \frac{d\delta k}{2\pi} e^{i (v(k_0)t -i x)\delta k+\frac{i}{2}\partial _k^2\varepsilon(k_0)t\delta k^2 +\frac{i}{6}\partial _k^2\varepsilon(k_0)t\delta k^3 +...},
\label{eq:saddle}
\eea
where $v(k_0)$ is the group velocity at $k_0$. We determine the butterfly velocity by requiring $u(v_Bt, t)$ changing slowly in the limit $t\rightarrow\infty$, which requires that both the first order term and the second order term vanish. This condition gives that the butterfly velocity $v_B$ is the maximal group velocity.

\textit{$t^{1/3} $wavefront spreading-} We move away from the wavefront a little bit by setting $v_B t-x <0$. In this case, the first order term is always nonzero. We keep up to the third order term and obtain that
\be
u(x,t)=\frac{e^{i (\varepsilon(k_0)t-k_0 x)}}{(-\partial_k^3 \varepsilon(k_0)t/2)^{1/3}} A_i \left (\frac{x-v_Bt}{(-\partial_k^3 \varepsilon(k_0)t/2)^{1/3}} \right)
\ee
where $Ai(Z)$ is the Airy function. We note that here $\partial_k^3 \varepsilon(k_0)$ is negative since the group velocity is maximal at $k_0$. In the limit that $|\partial_k^3 \varepsilon(k_0)t/2|^{1/3}\ll x-v_Bt\ll x$, we can use the asymptotic form of the Airy function and obtain Eq.~\eqref{eq:free} with $\lambda=\frac{2}{3}|\partial_k^3 \varepsilon(k_0)/2|^{-1/2}$

\textit{1/t late-time decay-} When $v_B t > x$, the saddle point exists and determined by the equation $v_0(k)=x/t$. In the limit that $t\gg x$, the saddle point occurs at where $v_0(k)=0$, and the second order term does not vanish. Then the integral in Eq.~\eqref{eq:saddle} gives
\be
u(x,t)\sim e^{i (\varepsilon(k_0)t-k_0 x)} \frac{1}{(m t)^{1/2}},
\ee
indicating that in the late limit, the squared commutator features a $1/t$ decaying tail.

\begin{figure}
\includegraphics[height=0.45\columnwidth, width=0.45\columnwidth]
{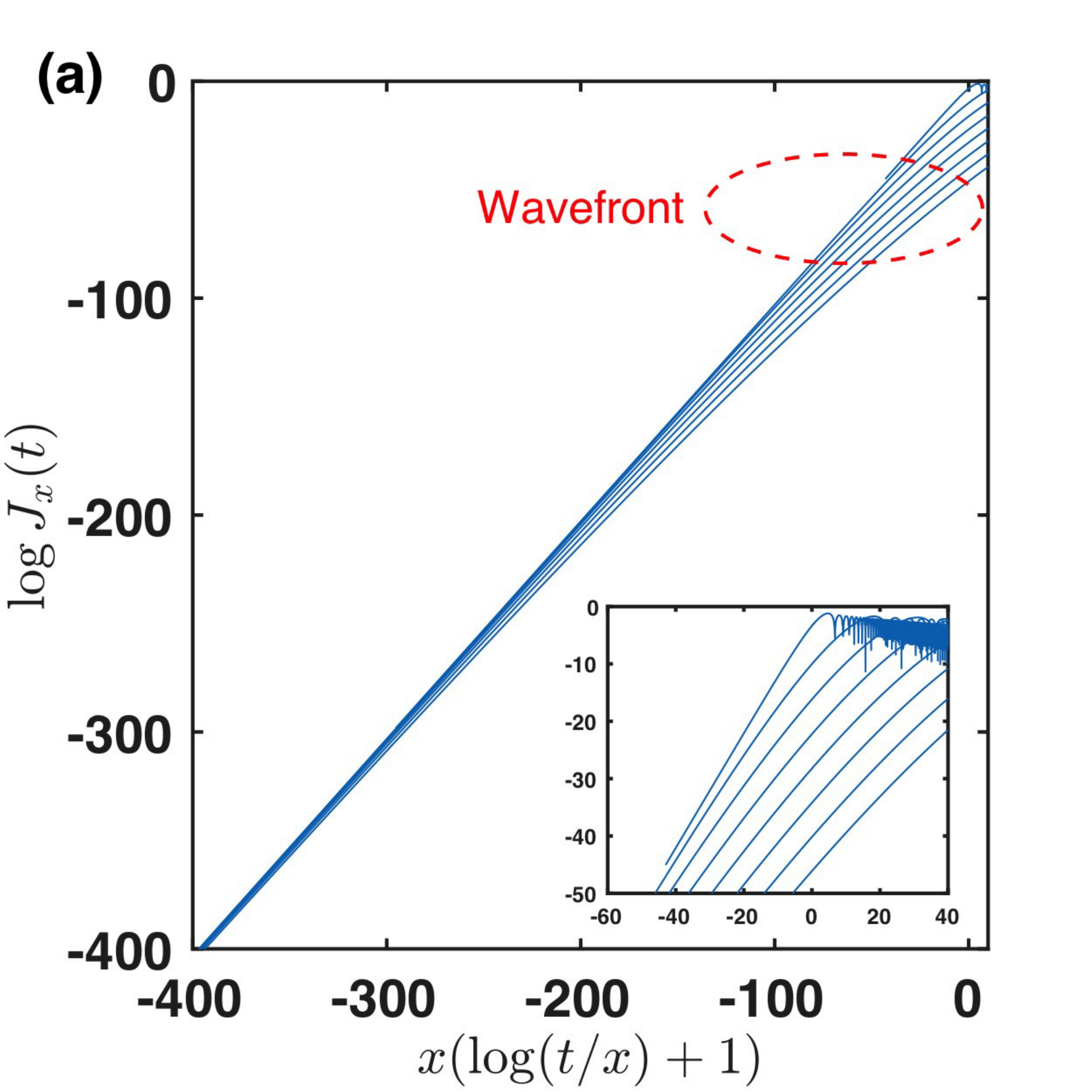}
\includegraphics[height=0.45\columnwidth, width=0.45\columnwidth]
{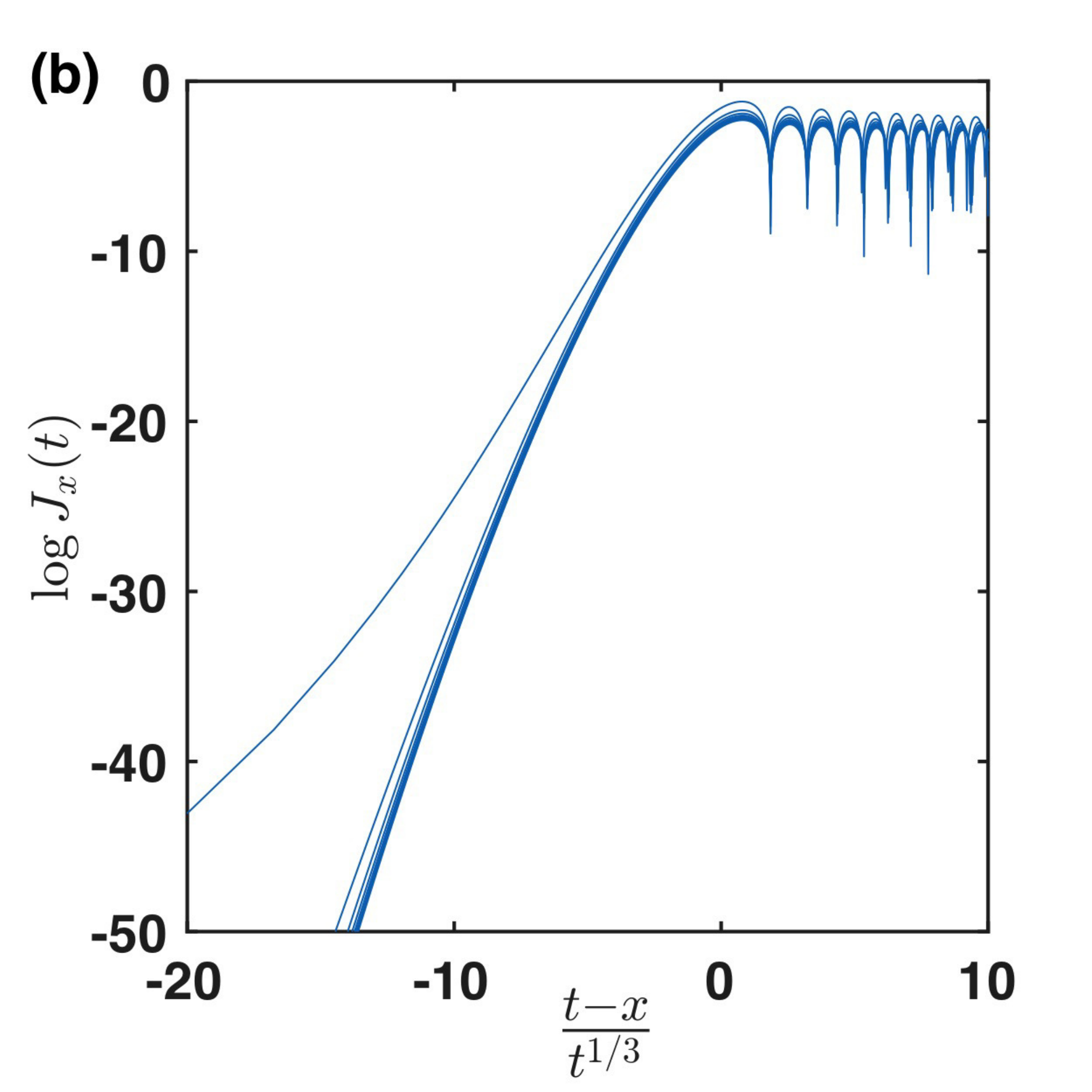}
\caption{Distinguishing the early-time behavior and early-growth behavior of the Bessel function. (a) In the early-time, the Bessel function $J_x(t)\sim \frac{(t/2)^x}{x!}$, which breaks down at the wavefront. (b) At the wavefront, the Bessel function collapses to the function with argument $(t-x)/t^{1/3}$.  Bessel functions of order from 10 to 300 are shown. }
\label{fig:bessel}
\end{figure}

\textit{Examples-}Now we provide some concrete examples. The simplest example is the 1D tight-binding model with a $\cos(k)$ band. In this case, we have $c_x(t)=\sum\limits_y  J_{x-y}(2t) c_y$,
and the squared commutator of the density operators is
\be
C(x,t)\sim J_{x}^2(2t).
\ee
Another example is the transverse-field Ising model at the critical point. In Appendix~\ref{appsc:heisenberg}, we show that
\be
C(x,t)\sim J_{2x}^2(4Jt)
\ee
Therefore, one can explicitly check all the above general properties with Bessel function. In particular, we note that from the continuum limit of the recurrence relation of the Bessel function, one can obtain that $\lim J_{x}(x+\delta t)\sim Ai(\delta t/x^{1/3})$, consistent with the $1/3$ wavefront spreading shown above. The Bessel function is also a nice example showing the crossover from the early-time region, dominated by the perturbative growth $\frac{(t/2)^x}{x!}$ which is simply its first term in the Taylor series,  to the non-trivial wavefront spreading form in the early growth region. The collapsing of functions with the two different forms is shown in Fig. \ref{fig:bessel}. We speculate that the crossover happens in generic many-body systems but with different value of $p$ near the wavefront.

As a final remark of this section, we note that in superconductors and multi-band systems, because of the interference between the different single-particle modes, the phase factor $e^{i \epsilon(k_0) t-k_0 x}$ in Eq.~\eqref{eq:saddle}  enters the squared commutator as a non-universal oscillating prefactor, which one can explicitly see in the transverse-field Ising model away from the critical point.

\section{Up bound of the operator entanglement entropy for the transverse-field model}
\label{appsc:up_bound}
In this section, we prove that the operator entanglement of a density operator in 1D fermion systems is up bounded by $\log(4)$, and so is $X(t)$ under Heisenberg dynamics of Eq.~\eqref{eq:ham}.

We cut the system at bond x and divide it into two subregions A and B. The Heisenberg dynamics of a local density operator $n$ takes the following form
\bea
n_0(t)&=c_0(t)^\dagger c_0(t)\\ &=\sum \left ( u_i (t) c^\dagger_i+v_i(t) c_i \right) \left ( v^*_j(t)  c^\dagger_j+u^*_j(t)  c_j\right) \\
&=\sum\limits_{i, j\in A/B} \left ( u_i (t) c^\dagger_i+v_i(t) c_i \right) \left ( v^*_j(t)  c^\dagger_j+u^*_j(t)  c_j \right)\\
&+\sum\limits_{i\in A, j\in B} \left ( u_i (t) c^\dagger_i+v_i(t) c_i \right) \left ( v^*_j(t)  c^\dagger_j+u^*_j(t)  c_j \right)\\
&-\sum\limits_{i\in A, j\in B}  \left ( v^*_i(t)  c^\dagger_i+u^*_i(t)  c_i \right)\left ( u_j (t) c^\dagger_j+v_j(t) c_j \right)
\eea
At $t=0$, $u_i=\delta_{i0}$ and $v_i=0$.
After Jordan-Wagner transformation, we have,
\bea
n_0(t)&=\sum\limits_{i, j\in A/B} \left (\prod\limits_{i<k<j} (-2S^z_k)\right ) \\
&\ \ \ \ \ \ \left ( u_i (t) S^+_i+v_i(t) S^-_i \right) \left ( v^*_j(t)  S^+_j+u^*_j(t)  S^-_j \right)\\
&+\sum\limits_{i\in A, j\in B} \left (\prod\limits_{i<k<x} (-2S^z_k)\right ) \left ( u_i (t) S^+_i+v_i(t) S^-_i \right) \\ &\ \ \ \ \ \ \ \ \ \ \ \ \ \left (\prod\limits_{x<k<j} (-2S^z_k)\right ) \left ( v^*_j(t)  S^+_j+u^*_j(t)  S^-_j \right)\\
&-\sum\limits_{i\in A, j\in B}  \left (\prod\limits_{i<k<x} (-2S^z_k)\right ) \left ( v^*_i(t)  S^+_i+u^*_i(t)  S^-_i \right)\\
& \ \ \ \ \ \ \ \ \ \ \ \ \ \left (\prod\limits_{x<k<L} (-2S^z_k)\right )\left ( u_j (t) S^+_j+v_j(t) S^-_j \right)
\eea
At the bond x, $n(t)$ can be then casted into a form of the matrix-product operator with bond dimension 4,
\be
n_0(t)=\left ( I_A, d_A, S_{A}, S_{A}^\dagger \right )\otimes \left ( d_B, I_B, S_{B}^\dagger, -S_{B} \right )^T
\ee
where
\bea
d_{A/B}=&\sum\limits_{i, j\in A/B} \left (\prod\limits_{i<k<j} (-2S^z_k)\right )\\
&\ \ \ \ \ \ \ \ \ \ \left ( u_i (t) S^+_i+v_i(t) S^-_i \right) \left ( v^*_j(t)  S^+_j+u^*_j(t)  S^-_j \right) \\
S_{A}=&\sum\limits_{i\in A} \left (\prod\limits_{i<k<x} (-2S^z_k)\right ) \left ( u_i (t) S^+_i+v_i(t) S^-_i \right)\\
S_{B}=&\sum\limits_{i\in B} \left (\prod\limits_{x<k<j} (-2S^z_k)\right )\left ( u_i (t) S^+_j+v_i(t) S^-_j \right)
\eea
Thus we have proved that the entanglement entropy of $n(t)$ is bounded by $\log 4$ for any cut.
This is also true for the operator $X(t)$ under dynamics governed by Eq.~\eqref{eq:ham} with $h_z=0$, since it is $2 n(t)-1$ after Jordan-Wigner transformation and has the following MPO form,
\be
X(t)=2\left ( I_A, d_A, S_{A}, S_{A}^\dagger \right )\otimes \left ( d_B-\frac{1}{2}I_B, I_B, S_{B}^\dagger, -S_{B} \right )^T
\ee

\section{Benchmarking MPO with exact diagonalization}
\label{appsc:benchmark}
\begin{figure}
\includegraphics[height=0.5\columnwidth, width=0.5\columnwidth]
{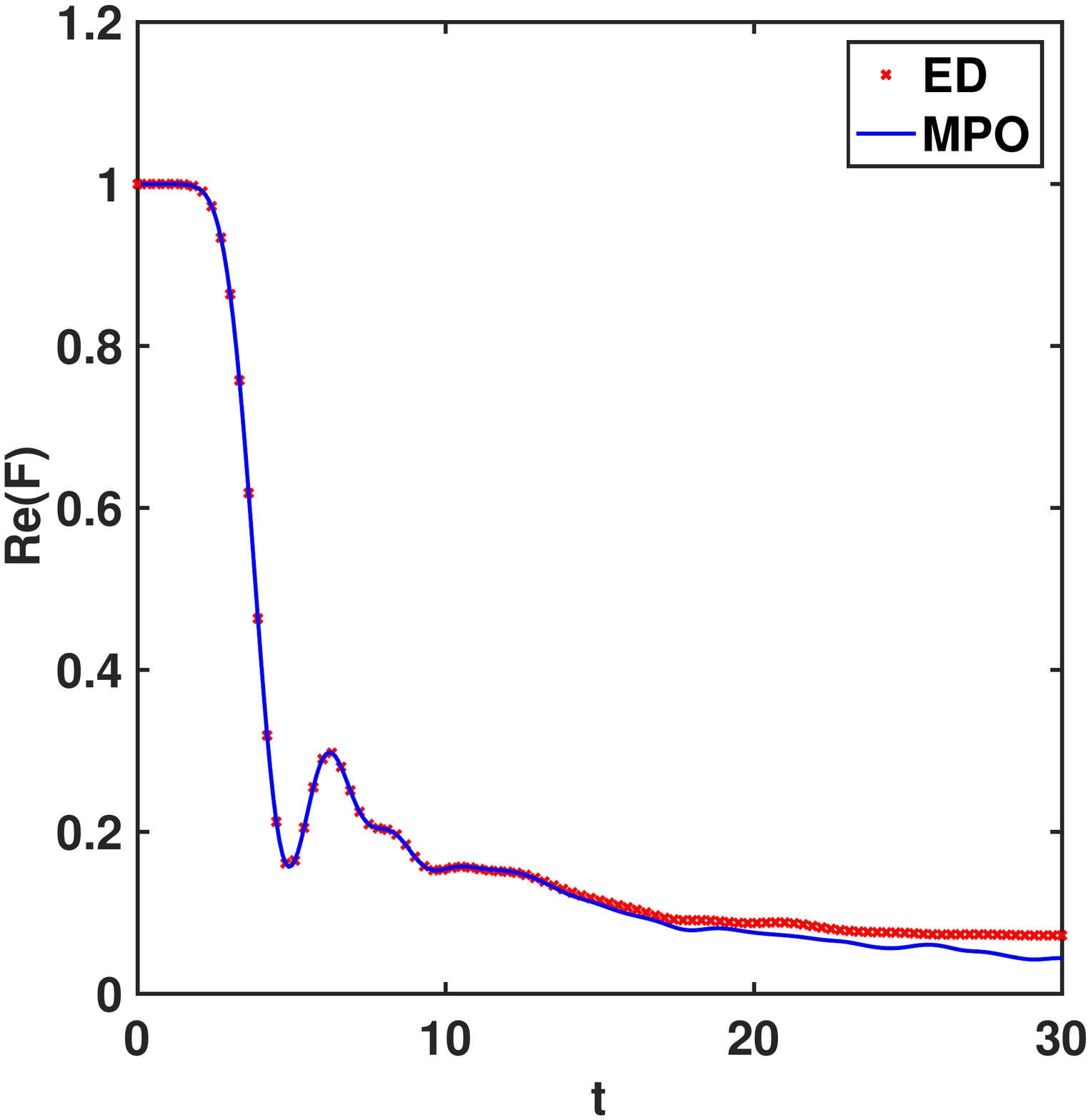}
\includegraphics[height=0.5\columnwidth, width=0.5\columnwidth]
{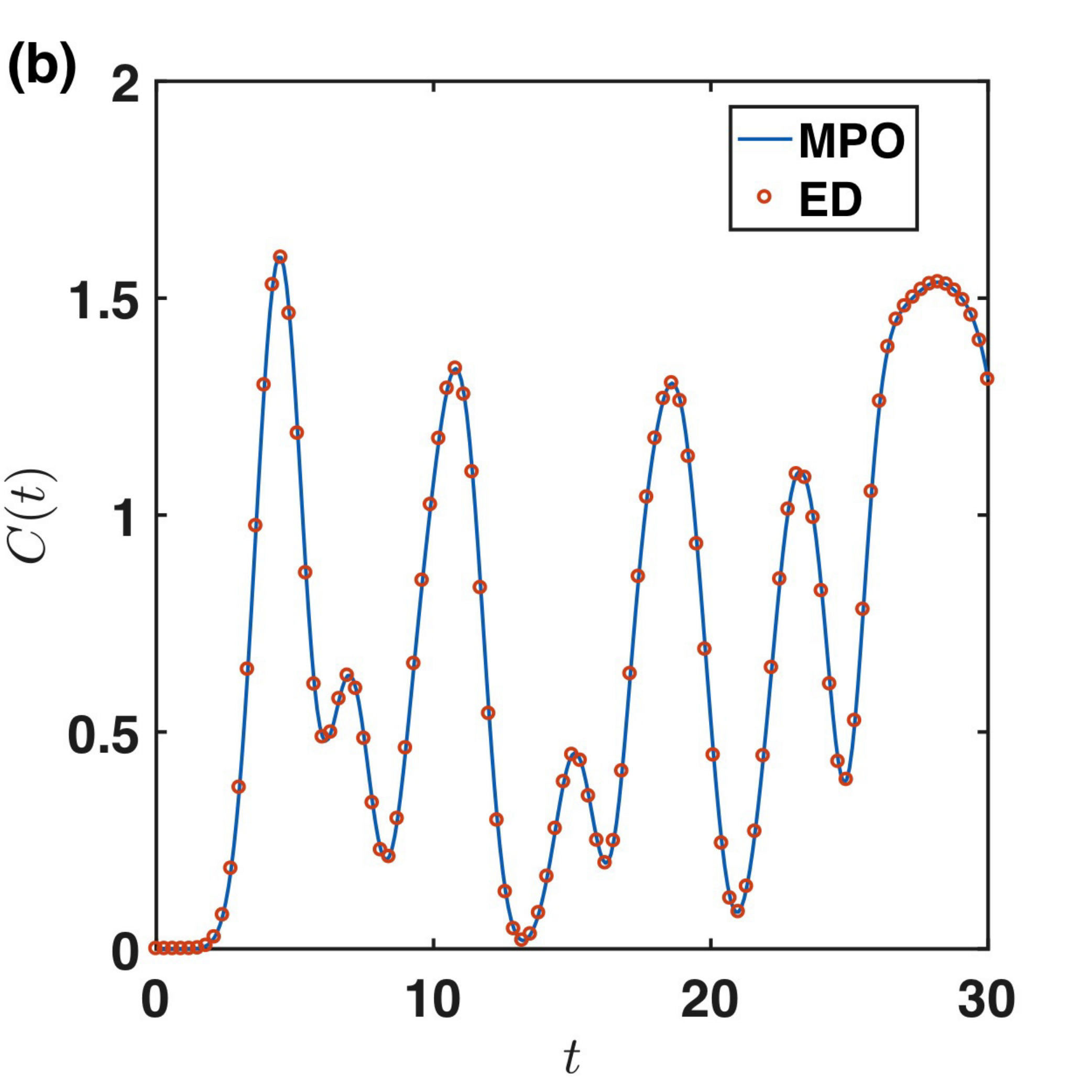}
\includegraphics[height=0.5\columnwidth, width=0.5\columnwidth]
{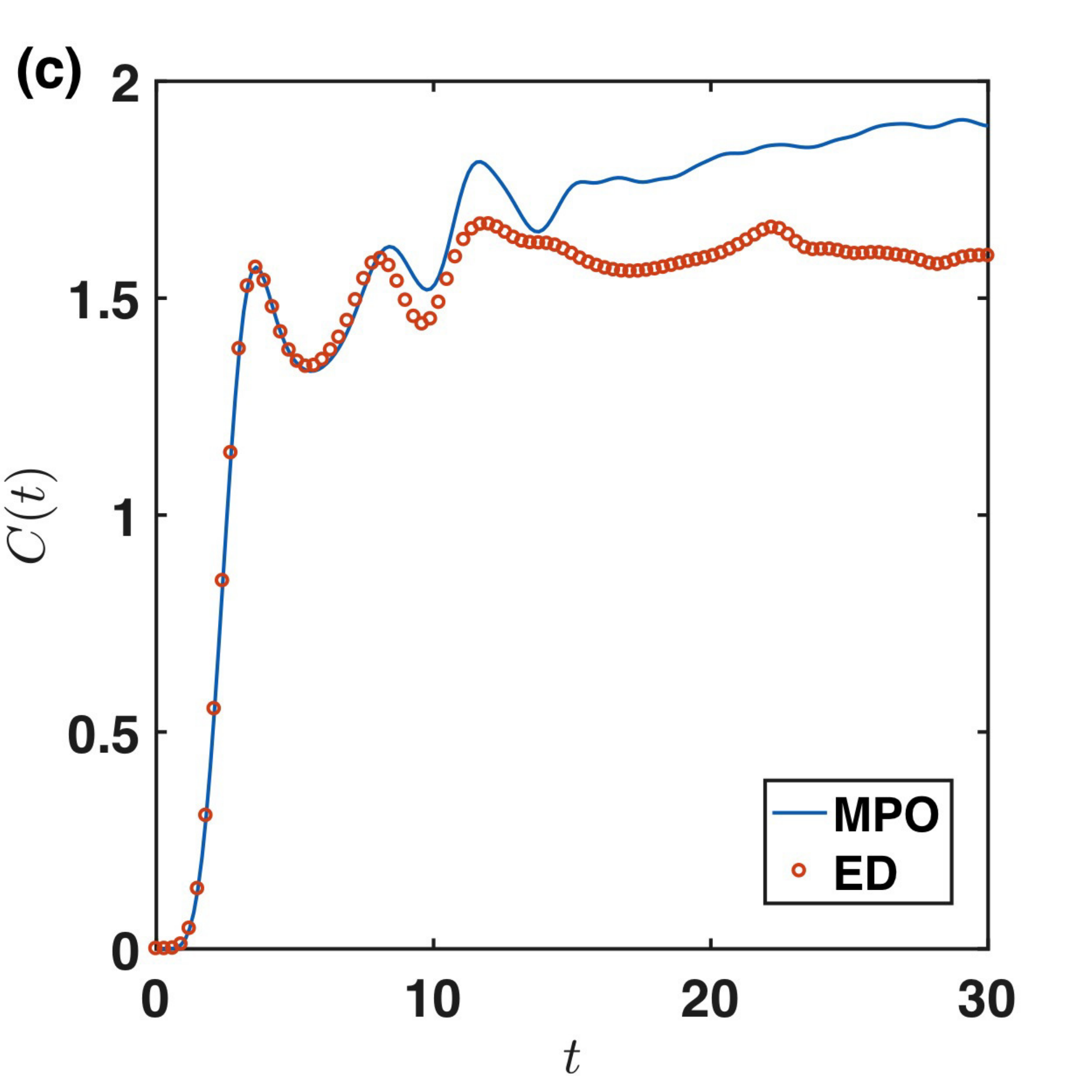}
\caption{Benchmarking the MPO results with the exact diagonalization on (a) mixed-field Ising model, (b) transverse-field Ising model and (c) Heisenberg model. In all three cases, in the early-growth region, the two methods generate identical results. }
\label{fig:benchmark}
\end{figure}

In this section, we benchmark the results obtained from MPO dynamics with ED on small system with $L=11$. The comparisons for chaotic mixed-field Ising model, transverse-field Ising model and the Heisenberg model are shown in Fig. \ref{fig:benchmark}. For all three cases, MPO agrees well with ED in the early growth region. But in the late-time region, the quality of the agreement varies: for the transverse-field model, MPO is able to produce the correct behavior of OTOC for arbitrarily long time because of the bounded operator entanglement; for the chaotic model, the MPO result is qualitatively correct; for the Heisenberg model, the late-time value of the squared commutator does not relax to 2 due to the integrability as shown by ED, but the MPO fails to reproduce that behavior. Designing an algorithm to correctly calculate the late behavior of scrambling will be an interesting future project.

\section{The early growth of OTOC for the transverse-field model and the Heisenberg model}
\label{appsc:heisenberg}
In this section, we present results on the behavior of OTOC in the early growth region for the transverse-field Ising models and the Heisenberg model.

The transverse-field model is quadratic in terms of the fermion operators after the Jordan-Wagner transformation, allowing us to calculate the squared commutator Eq.~\eqref{eq:cinfty} analytically. Under Heisenberg dynamics, the annihilation operator $c_x(t)$ takes the following form,
\be
c_x(t)=\sum_y u(x-y) c_y(t)+v(x-y) c_y^\dagger(t),
\ee
where
\bea
&u(x)=\int\limits_{-\pi}^{\pi}\frac{dk}{2\pi}\left (\cos (\varepsilon(k)t)+i \frac{\cos k+g}{\varepsilon(k)}\sin (\varepsilon(k)t) \right )\cos(kx)\\
&v(x)=-i\int\limits_{-\pi}^{\pi}\frac{dk}{2\pi}\frac{\sin k}{\varepsilon(k)}\sin (\varepsilon(k)t) \sin(kx)
\eea
with the dispersion $\epsilon(k)=\sqrt{1+g^2+2 g \cos k}$ and $g=h_x/J$. Here $t$ is in the unit of $2J$.
Then one can show that Eq.~\eqref{eq:cinfty}, the squared commutator of $X_i(t)$ and $X_j$, is
\be
C(x,t)=8 \left (|u(x)|^2+|v(x)|^2-(|u(x)|^2-|v(x)|^2)^2 \right )
\ee
At $g=1$, the function $u(x)$ and $v(x)$ can be evaluated analytically as $u(x)=(-1)^x(J_{2x}(2t)+iJ'_{2x}(2t))$  and $v(x)=i(-1)^x x/t J_{2x}(2t)$. We have checked our analytical calculation with our MPO dynamics, and they perfectly agree with each other for various values of $g$.

\begin{figure}
\includegraphics[height=0.45\columnwidth, width=0.45\columnwidth]
{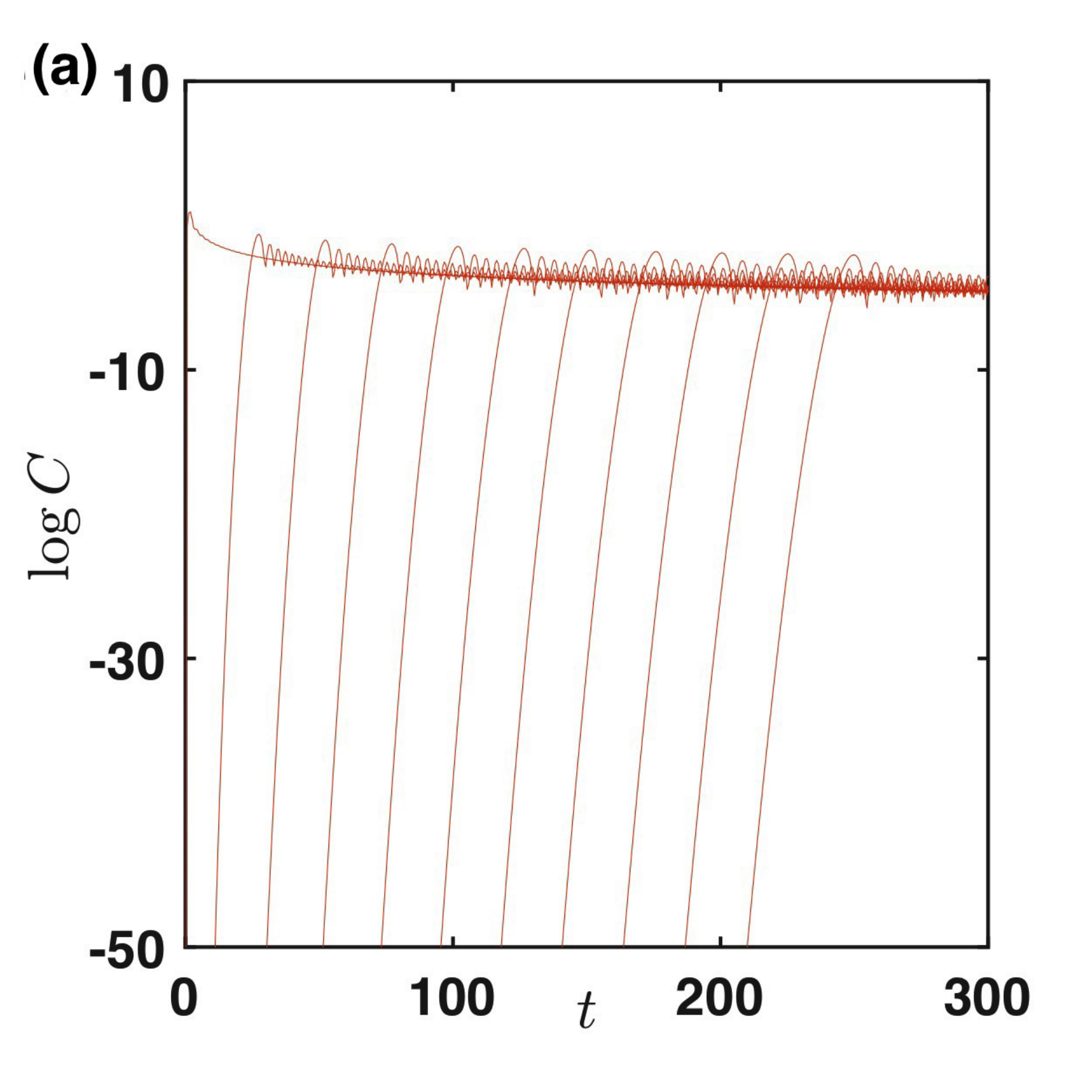}
\includegraphics[height=0.45\columnwidth, width=0.45\columnwidth]
{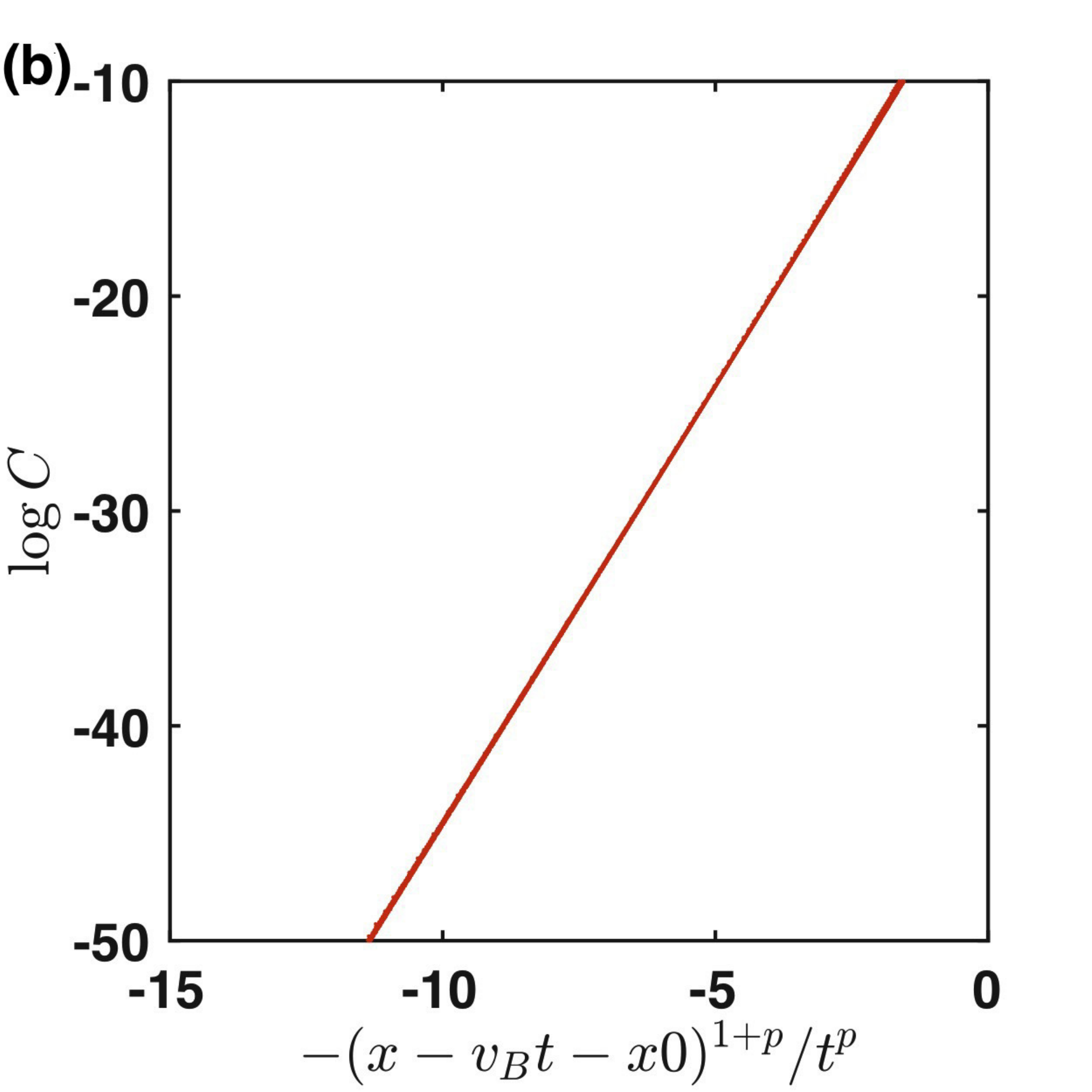}
\caption{(a) The early growth of the squared commutator of $X_i(t)$ and $X_j$ for the transverse-field model at the critical point. The curves with $j-i=0, 20, 40 ..., 200$ are shown.  (b) The early growth behavior is fitted into the form $e^{-\lambda (x-v_Bt-x_0)^{1+p}/t^p}$ with $\lambda=4.0$, $v_B=0.82$, $p=0.48$ and $x_0=0.6$. This indicates a sub-diffusively spreading wavefront. }
\label{fig:trsvs1}
\end{figure}

First, we verify the form Eq.~\eqref{eq:free} in the transverse-field model at the critical point where $g=1$. As shown in the Fig. \ref{fig:trsvs1}, the data fit of the early growth of scrambling clearly reveals that $p=1/2$ in this case.

\begin{figure}
\includegraphics[height=0.45\columnwidth, width=0.45\columnwidth]
{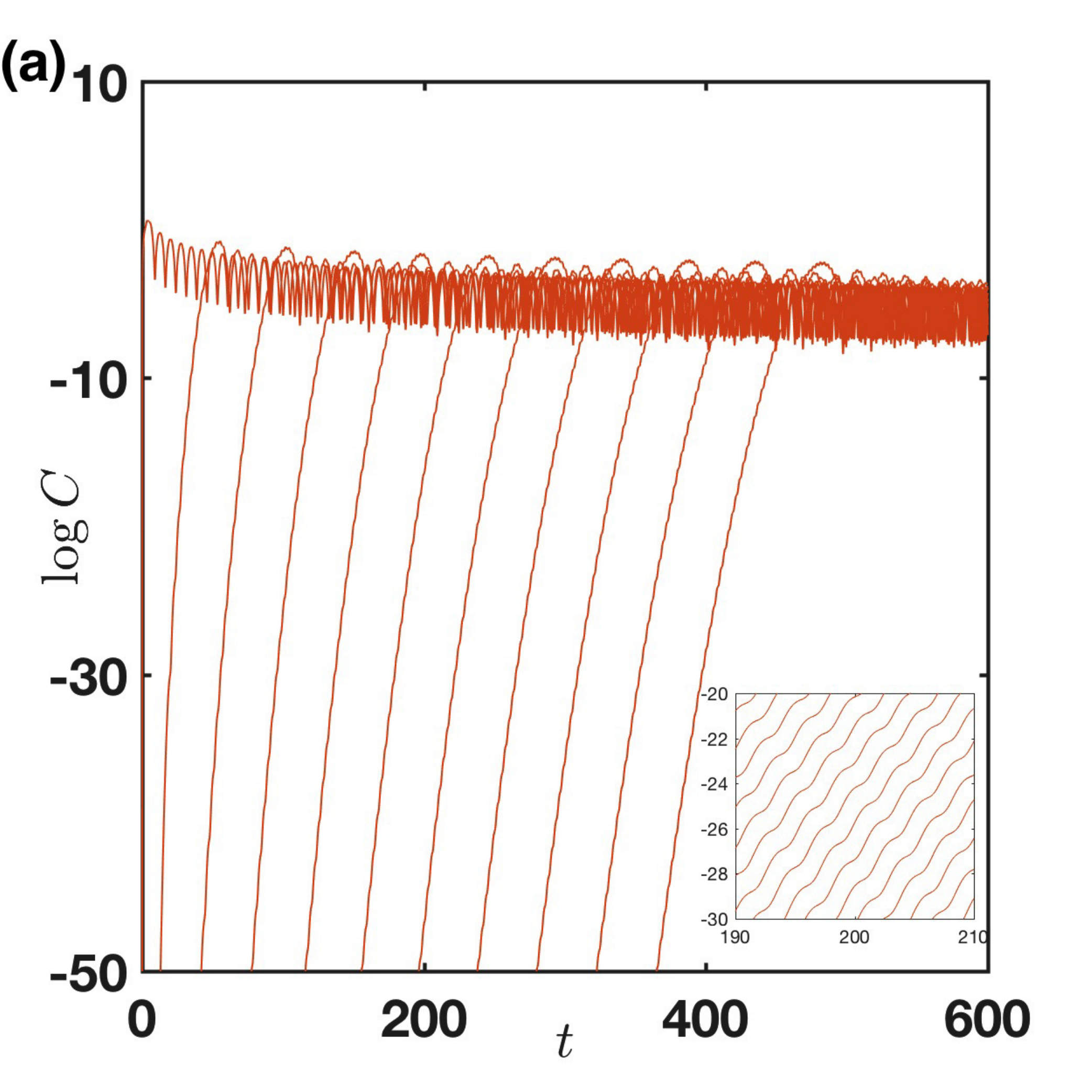}
\includegraphics[height=0.45\columnwidth, width=0.45\columnwidth]
{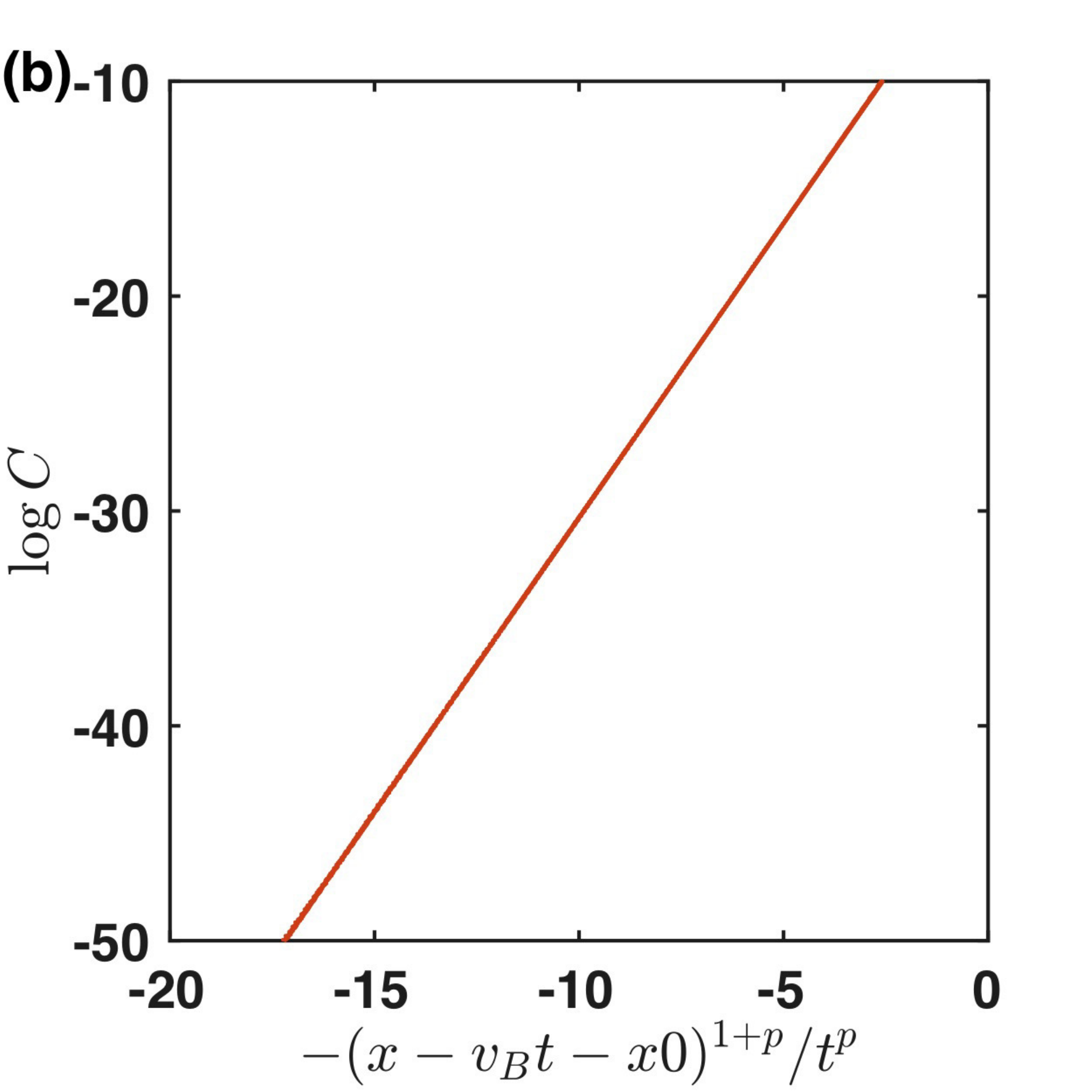}
\caption{(a) The early growth of the squared commutator of $X_i(t)$ and $X_j$ for the transverse-field model with $h_x=3J$. The curves with $j-i=0, 10, 20 ..., 200$ are shown. As shown in the inset, $C(x,t)$ picks up a non-universal oscillation depending on the energy gap. (b) After removing the oscillation and only fitting the smooth part of the data, we again obtain good agreement with the universal growth form Eq.~\eqref{eq:fitting_form} with $\lambda=2.7, v_B=0.42, p=0.482, x_0=-0.113.$}
\label{fig:trsvs3}
\end{figure}

Then we present the result of the case away from the critical point where $g=3$. As argued in the appendix, due to the interference between different single-particle mode, which are the Bogoliubov quasi-particles, the squared commutator carries a non-universal oscillating part, shown in Fig. \ref{fig:trsvs3}. After removing the oscillation and only fitting the smooth part of the data, we again get agreement between the squared commutator and the growth form Eq.~\eqref{eq:free}.

Finally, we present the early growth of scrambling in the Heisenberg chain. Similar to the case of the mixed-field Ising demonstrated in the main text, MPOs with $\chi=4$ and $\chi=32$ give almost identical results, which fit nicely into the universal growth form Eq.~\eqref{eq:fitting_form}.

\begin{figure}
\includegraphics[height=0.45\columnwidth, width=0.45\columnwidth]
{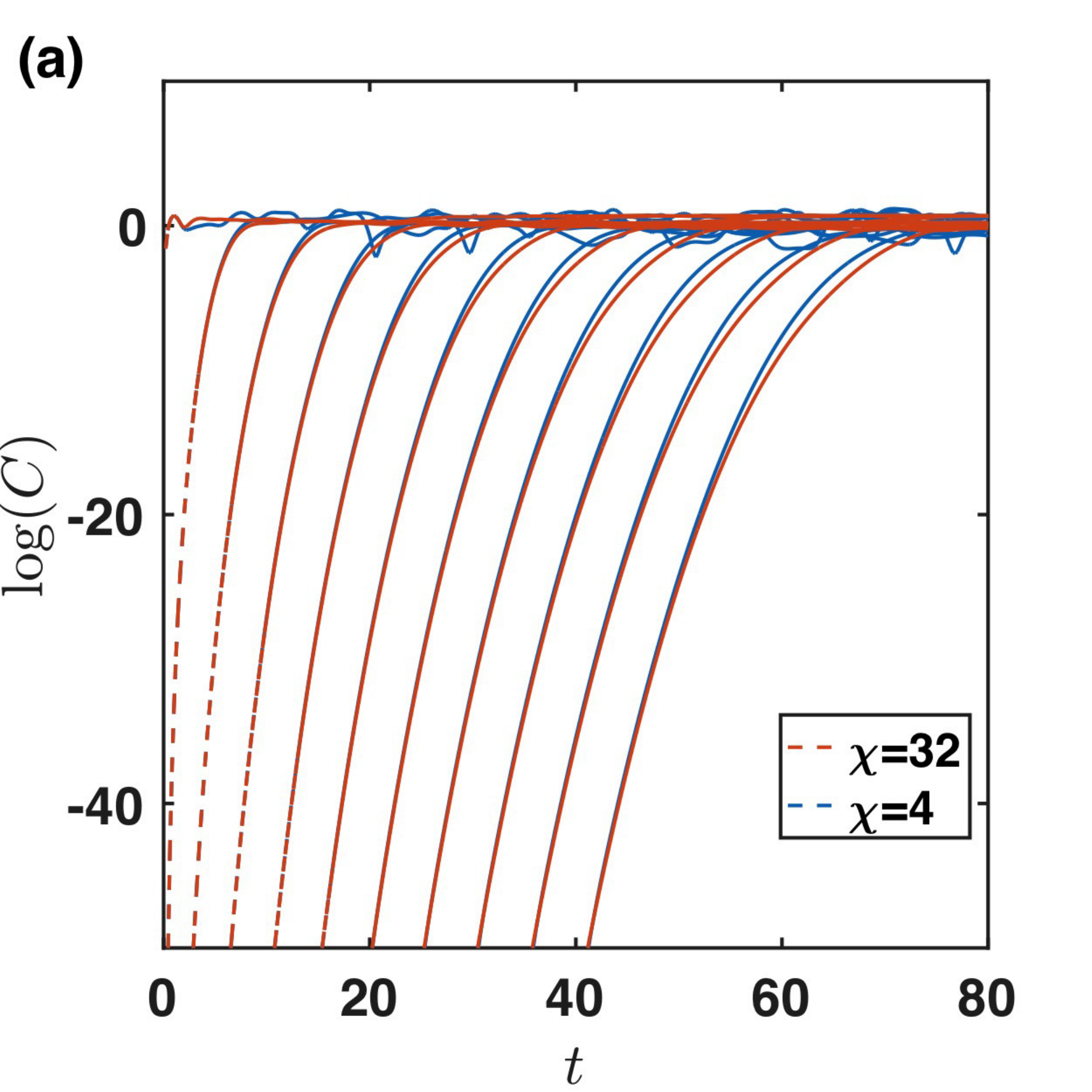}
\includegraphics[height=0.45\columnwidth, width=0.45\columnwidth]
{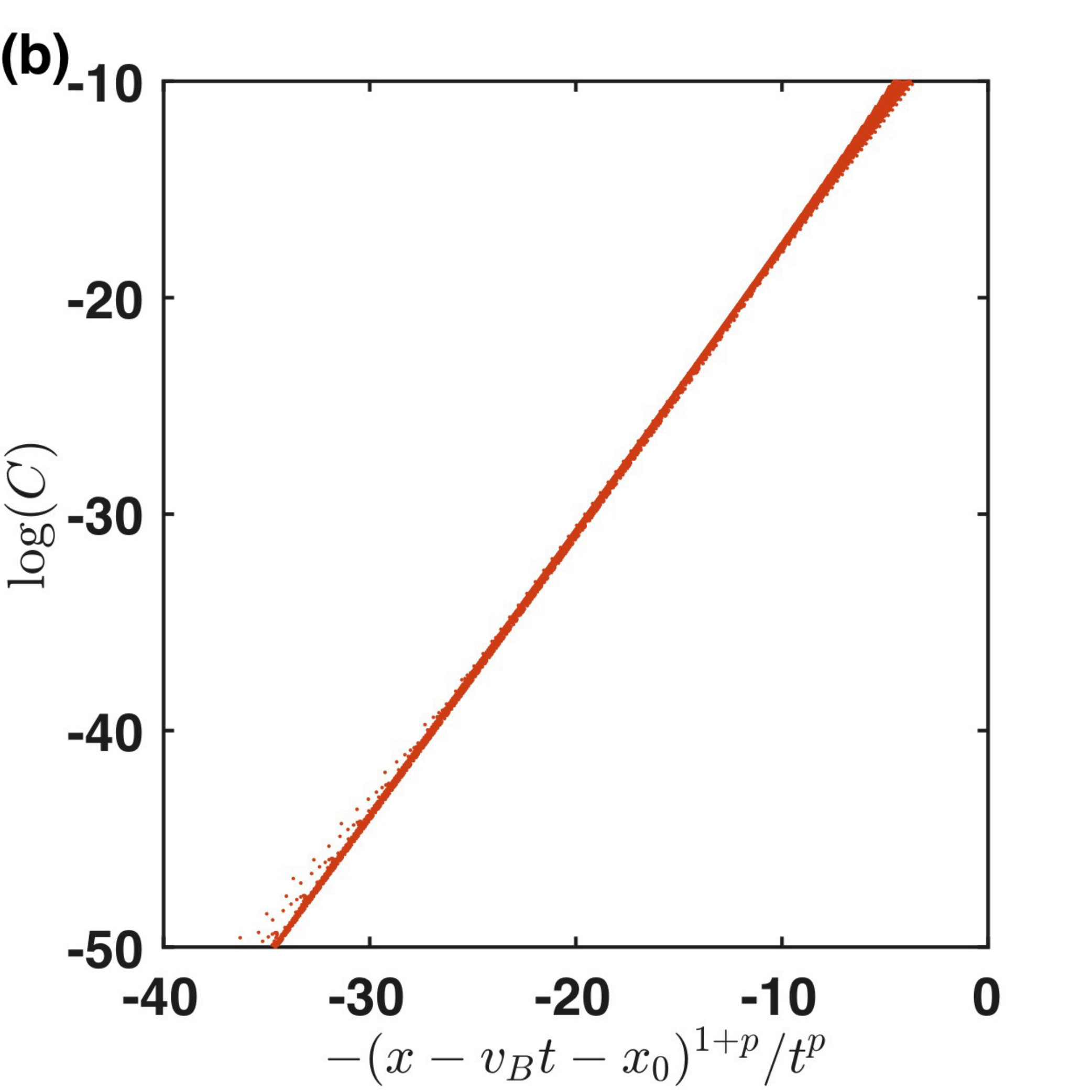}
\caption{(a) The early growth of the squared commutator of $X_i(t)$ and $X_j$ for the Heisenberg model. The curves with $j-i=0, 10, 20 ..., 100$ are shown.  (b) The early growth behavior is fitted into the form $e^{-\lambda (x-v_Bt-x_0)^{1+p}/t^p}$ with $\lambda=1.316$, $v_B=1.514$, $p=0.54$ and $x_0=1.79$. This indicates a sub-diffusively spreading wavefront. }
\label{fig:heisenberg}
\end{figure}

\end{document}